\documentclass[aps,prb,reprint,superscriptaddress,10pt,longbibliography]{revtex4-2}
\usepackage[colorlinks,bookmarks=false,citecolor=blue,linkcolor=blue,urlcolor=blue]{hyperref}
\usepackage{amsmath, amssymb, graphicx, color}
\bibpunct{[}{]}{,}{n}{}{} 

\newcommand{\Imm}{\mathop{\rm Im}\nolimits}
\newcommand{\sgn}{\mathop{\rm sgn}\nolimits}

\begin{document}

\title{Evolution of the Andreev bands in the 
half-filled superconducting periodic Anderson model}

\author{Vladislav Pokorn\'y}
\email{pokornyv@fzu.cz}
\affiliation{Institute of Physics, Czech Academy of Sciences, 
Na Slovance 2, CZ-18221 Praha 8, Czech Republic}

\author{Panch Ram}
\email{panchram@karlov.mff.cuni.cz}
\affiliation{Department of Condensed Matter Physics, Faculty of Mathematics and Physics, 
Charles University, Ke Karlovu 5, CZ-12116  Praha 2, Czech Republic}

\date{\today}

\begin{abstract}

We employ the periodic Anderson model with superconducting 
correlations in the conduction band at half filling to study 
the behavior of the in-gap bands in a heterostructure consisting 
of a molecular layer deposited on the surface of a conventional 
superconductor. We use the dynamical mean-field theory to map the 
lattice model on the superconducting single impurity model with 
self-consistently determined bath and use the continuous-time 
hybridization expansion (CT-HYB) quantum Monte Carlo and the 
iterative perturbation theory (IPT) as solvers for the impurity 
problem. We present phase diagrams for square and triangular
lattice that both show two superconducting phases that differ by 
the sign of the induced pairing, in analogy to the $0$ and $\pi$ 
phases of the superconducting single impurity Anderson model and 
discuss the evolution of the spectral function in the vicinity of 
the transition. We also discuss the failure of the IPT for 
superconducting models with spinful ground state and the behavior 
of the average expansion order of the CT-HYB simulation.
\end{abstract}


\maketitle


\section{Introduction \label{Sec:Intro}}
Individual magnetic impurities embedded in a metallic host give rise to the
Kondo effect that describes the screening of the local moment
by the conduction electrons, resulting in a many-body, non-magnetic 
singlet ground state~\cite{Hewson-1993}. A footprint of this effect is a narrow resonance 
in the density of states (DOS) and its width is connected with the energy scale 
$k_BT_K$, where $T_K$ is the so-called
Kondo temperature, which quantifies the exchange interaction between the impurity and the bath.
The same effect takes place when a spinful 
atom or molecule is adsorbed on a surface of a metal~\cite{Zhao-2005}. In both cases,
the physics is well captured by the single impurity Anderson model (SIAM)
that describes a single energy level hybridized with a bath of conduction electrons.

A different situation arises if the metallic host is replaced by 
a superconductor. Here the conduction electrons with antiparallel spins from the 
vicinity of the Fermi surface form a singlet bound state as described by the BCS
theory. As a result, a gap is opened at the Fermi energy and the screening
of the local moment of the impurity can be incomplete. Furthermore,
Andreev reflection of the Cooper pairs off the impurity results in
the presence of a set of in-gap states known as Andreev bound states 
(ABS)~\cite{Balatsky-2006}. In the case of imperfect screening the
strong on-site Coulomb repulsion can drive the system to a magnetic, doublet ground state.
This competition between the screening 
and the superconductivity can be quantitatively described 
by the ratio of the two relevant energy scales, the Kondo temperature $k_BT_K$
and the superconducting gap $\Delta$. For $k_BT_K\gg\Delta$ the system
is in non-magnetic state, while for $k_BT_K\ll\Delta$ the ground 
state is magnetic. If the two energy scales become comparable, 
the system undergoes a transition marked by the crossing of the lowest-lying 
ABS at the Fermi energy as a result of the change of the many-body ground state. 
This is an example of an impurity quantum phase transition~\cite{Vojta-2006} best 
known as the $0-\pi$ transition and 
it is well captured by the superconducting impurity Anderson model 
(SCIAM)~\cite{Luitz-2012}.
It was experimentally observed in a number of setups including
semiconducting nanowires and nanotubes proximitized to superconducting 
electrodes~\cite{Cleuziou-2006,Jorgensen-2007,Wernsdorfer-2010} as well 
as single atoms or molecules adsorbed on superconducting 
surfaces~\cite{Franke-2011,Odobesko-2020}.

When the concentration of the impurities is large, the electron hopping,
either direct or via the conduction band, gives rise to the dispersion of 
the local impurity level and a correlated impurity band of finite width is formed. 
Such a system with a metallic host can be described by the periodic Anderson model (PAM). 
This lattice model was originally developed to study the physics of heavy fermion 
compounds like $\mathrm{SmB_6}$~\cite{Menth-1969} or $\mathrm{YbB_{12}}$~\cite{Iga-1988}
and its physics is much richer than of the SIAM. Its phase diagram consist of
a Kondo insulator (KI), metallic and Mott insulator phase~\cite{Logan-2016},
and a variety of magnetic phases~\cite{Vekic-1995,Ram-2018}. 
It was also used to study the possibility of unconventional superconductivity mediated by
spin fluctuations~\cite{Bodensiek-2013,Wu-2015} and magnetic quantum oscillations 
under the orbital response to the magnetic field~\cite{Ram-2017,Ram-2019}.
PAM with attractive interactions in the impurity band was used to simulate 
the superfluid Bose-Einstein condensate in ultracold 
atoms~\cite{Anderson-1995,Koga-2010,Koga-2011} and to study the local 
magnetic moment formation in presence of superconducting 
correlations~\cite{Araujo-2001}.

A direct generalization of the SCIAM to a lattice model is the 
superconducting PAM (SCPAM) which describes a single correlated electron 
orbital at each lattice site hybridized with a band of conduction electrons 
with local superconducting pairing. A very successful approach to study 
this class of systems is the dynamical mean-field theory (DMFT)~\cite{Georges-1996}.
This method  maps the lattice model to an impurity model with 
a self-consistently determined bath. SCPAM was studied using DMFT by 
Luitz and Assaad~\cite{Luitz-2010}. The authors used the 
continuous time interaction-expansion (CT-INT) quantum Monte Carlo (QMC) as the solver for 
the impurity problem. Results show that the $0-\pi$ transition of SCIAM is inherited to SCPAM
as a first order transition. 
Oei and Tanaskovi\'{c}~\cite{Oei-2020} later used SCPAM to study 
the effect of magnetic impurities on a bulk $s$-wave superconductor using
hybridization-expansion (CT-HYB) QMC and the dual fermion method in 
attempt to explain the reentrant behavior of the superconducting phase 
in certain classes of unconventional superconductors.

The SCPAM can be also employed to describe an atomic or molecular layer 
deposited on the surface of a superconductor where the interplay of 
superconductivity and magnetism gives rise to a number of interesting phenomena.
An example of such a setup is a van der Waals heterostructure consisting of thin
superconducting layer coated with a layer of transition metal (e.g., Mn, Fe, Co)
phtalocyanine molecules~\cite{Yoshizawa-2017,Uchihashi-2019}
in which the effect of the magnetic state of the molecule on the
superconducting properties of the substrate was studied. Such molecules 
form a superlattice of impurities that may or 
may not be commensurate with the structure of the surface.
In the ideal case of commensurate order,
such superlattices can be effectively treated using single-site DMFT
as the larger distance between the impurity sites weakens the spatial correlations,
making DMFT a better approximation than for bulk systems. 
Another example of such system is the hydrogenated superconducting 
graphene~\cite{Lado-2016} where individual hydrogens 
interact ferromagnetically when deposited 
on the graphene and can give rise to in-gap Andreev bands.
Also, recent experiments involving superconducting 
boron-doped diamond coated with ferromagnetic 
hydrogen monolayer~\cite{Zhang-2017,Zhang-2020} show the existence of Andreev bands 
and regions of high zero-bias conductance. 

The paper is organized as follows. In Sec.~\ref{Sec:SCPAM} we introduce
the Hamiltonian of the SCPAM and rewrite it in the Nambu formalism.
In Sec.~\ref{Sec:DMFT} we derive the DMFT equations for the superconducting 
system and introduce two methods of solving the impurity problem, CT-HYB
and the iterative perturbation theory (IPT),
and discuss their pros and cons for the given problem.
In Sec.~\ref{Sec:Results} we present the main results: 
In Sec.~\ref{SSec:Phasediag} we present the phase diagrams 
of SCPAM at constant temperature and half filling on square 
and triangular lattices and we discuss the character of the 
two emerging superconducting phases. 
The transition between the two superconducting phases is further
illustrated in Sec.~\ref{SSec:Spectra}
on the behavior of the spectral functions calculated using 
IPT. We close this section by Sec.~\ref{SSec:ExpOrder} where we look into 
the behavior of the average expansion order of the CT-HYB simulation and what
information can be extracted from its behavior. We also use this
quantity to discuss the temperature dependence of the phase boundaries.
The main points are summarized in Sec.~\ref{Sec:Conclusions}.
To make the paper more self-contained, in Appendix~\ref{App:DOSzero} 
we present analytic formulas for the bare local Green functions 
and in Appendices~\ref{App:Nonint} and~\ref{App:Nonint2} we summarize the 
basic properties of the non-interacting model.

\section{SCPAM\label{Sec:SCPAM}}

\subsection{Hamiltonian \label{SSec:Ham}}
The Hamiltonian of the SCPAM
describes a band of conduction electrons with local superconducting pairing
that hybridizes with a non-dispersive, correlated electron orbital at each 
lattice site $i=1,\ldots,N$ and reads
\begin{equation}
\label{Eq:HamPAM}
\mathcal{H}=\mathcal{H}_{d}+\mathcal{H}_{c}+\mathcal{H}_{\mathrm{hyb}}.
\end{equation}
The Hamiltonian describing the correlated sites reads
\begin{equation}
\mathcal{H}_{d}=\sum_{i\sigma}\varepsilon_{\sigma}
d_{i\sigma}^\dag d_{i\sigma}^{\phantom{\dag}}
+U\sum_i d_{i\uparrow}^\dag d_{i\uparrow}^{\phantom{\dag}}
d_{i\downarrow}^\dag d_{i\downarrow}^{\phantom{\dag}},
\end{equation}
where $d_{i\sigma}^\dag$ creates an electron
at site $i$ with spin 
$\sigma\in\{\uparrow,\downarrow\}$ and energy 
$\varepsilon_{\sigma}=\varepsilon-\mu-\sigma h$ where
$h$ is the local magnetic field,
$\mu$ is the chemical potential, and $U$ is the local on-site Coulomb repulsion. 
The attractive interaction in the conduction band is treated on the static 
mean-field level. Its Hamiltonian reads
\begin{equation}
\mathcal{H}_{c}=\sum_{\mathbf{k}\sigma}\varepsilon_{\mathbf{k}\sigma}
c_{\mathbf{k}\sigma}^\dag c_{\mathbf{k}\sigma}^{\phantom{\dag}}
-\Delta\sum_{\mathbf{k}}(c_{\mathbf{k}\uparrow}^\dag 
c_{\mathbf{\bar{k}}\downarrow}^\dag + \mathrm{H.c.}).
\end{equation}
Here $c_{\mathbf{k}\sigma}^\dag$ creates an electron with spin $\sigma$ 
and energy 
$\varepsilon_{\mathbf{k}\sigma}=\varepsilon_{\mathbf{k}}-\mu-\sigma h$
and
\begin{equation}
\label{Eq:BCSdelta}
\Delta=\frac{g}{N}\sum_\mathbf{k}\langle 
c_{\mathbf{\bar{k}}\downarrow}c_{\mathbf{k}\uparrow}\rangle
\equiv \frac{g}{N}\sum_\mathbf{k}\langle 
c_{\mathbf{k}\uparrow}^\dag c_{\mathbf{\bar{k}}\downarrow}^\dag\rangle
\end{equation}
is the the BCS superconducting gap parameter where $g$ is the attractive, 
phonon-mediated interaction strength and we denoted 
$\mathbf{\bar{k}}=-\mathbf{k}$ to save space in the equations. 
Finally, the term describing the coupling between the two subsystems reads
\begin{equation}
\label{Eq:Hhyb}
\mathcal{H}_{\mathrm{hyb}}=-\sum_{\mathbf{k}\sigma}(V_{\mathbf{k}\sigma}
d_{\mathbf{k}\sigma}^\dag c_{\mathbf{k}\sigma}^{\phantom{\dag}} + \mathrm{H.c.}),
\end{equation}
where $V_{\mathbf{k}\sigma}$ denotes the hybridization matrix element,
\begin{equation}
d_{\mathbf{k}\sigma}^\dag=
\frac{1}{\sqrt{N}}\sum_{i}e^{-i\mathbf{k}\cdot\mathbf{r}_i}d_{i\sigma}^\dag,\quad
d_{\mathbf{k}\sigma}=
\frac{1}{\sqrt{N}}\sum_{i}e^{i\mathbf{k}\cdot\mathbf{r}_i}d_{i\sigma}
\end{equation}
and $\mathbf{r}_i$ is the position vector of the lattice site $i$.

It is convenient to use the Nambu formalism while dealing with 
superconducting Hamiltonians. We define the Nambu spinors
\begin{equation}
\label{Eq:NambuSpinors}
D_{\mathbf{k}}=\begin{pmatrix}
d_{\mathbf{k}\uparrow}^{\phantom{\dag}} \\[0.3em]
d_{\bar{\mathbf{k}}\downarrow}^{\dag} \\[0.3em]
\end{pmatrix},\quad
C_{\mathbf{k}}=\begin{pmatrix}
c_{\mathbf{k}\uparrow}^{\phantom{\dag}} \\[0.3em]
c_{\bar{\mathbf{k}}\downarrow}^{\dag}
\end{pmatrix},
\end{equation}
and matrices
\begin{equation}
\label{Eq:NambuMatrices}
\begin{aligned}
E_{c\mathbf{k}} &= 
\begin{pmatrix}
\varepsilon_{\mathbf{k}\uparrow} & -\Delta \\[0.3em]
-\Delta & -\varepsilon_{\mathbf{\bar{k}}\downarrow}
\end{pmatrix},\quad
E_d = 
\begin{pmatrix}
\varepsilon_{\uparrow} & 0 \\[0.3em]
0 & -\varepsilon_{\downarrow}
\end{pmatrix}, \\
V_{\mathbf{k}} &= 
\begin{pmatrix}
-V_{\mathbf{k}\uparrow} & 0 \\[0.3em]
0 & V_{\mathbf{\bar{k}}\downarrow}
\end{pmatrix},
\end{aligned}
\end{equation}
from which we construct the double-spinor $\psi_{\mathbf{k}}$ 
and the $4\times 4$ matrix $E_{\mathbf{k}}$,
\begin{equation}
\label{Eq:Nambu4}
\psi_{\mathbf{k}}=
\begin{pmatrix}
D_{\mathbf{k}} \\[0.3em]
C_{\mathbf{k}}
\end{pmatrix},\quad
E_{\mathbf{k}} =
\begin{pmatrix}
E_d & V_{\mathbf{k}} \\[0.3em]
V^\dag_{\mathbf{k}} & E_{c\mathbf{k}}
\end{pmatrix}.
\end{equation}
Hamiltonian in Eq.~\eqref{Eq:HamPAM} can be written in the form
\begin{equation}
\label{Eq:PAMhamNambu}
\mathcal{H}=
\sum_{\mathbf{k}}\psi_{\mathbf{k}}^\dag E_{\mathbf{k}} \psi_{\mathbf{k}}
+U\sum_i d_{i\uparrow}^\dag d_{i\uparrow}^{\phantom{\dag}}
d_{i\downarrow}^\dag d_{i\downarrow}^{\phantom{\dag}}.
\end{equation}

The structure of the underlying lattice is fully encoded in the 
dispersion relation $\varepsilon_{\mathbf{k}}$. Motivated by the 
experiment concerning superconducting boron-doped diamond~\cite{Zhang-2020}, 
we consider two types of lattices, square and triangular,
simulating a lattice of impurities on the (100) and (111) diamond surfaces. 
The square lattice is described by
\begin{equation}
\label{Eq:dispSq}
\varepsilon_{\mathbf{k}\square}=-2t[\cos(ak_x)+\cos(ak_y)],
\end{equation}
where $t$ is the nearest-neighbor hopping amplitude, $a$ is the lattice 
constant and we set $\hbar=1$ for simplicity.
The triangular lattice is described by
\begin{equation}
\label{Eq:dispTr}
\varepsilon_{\mathbf{k}\triangle}=
-2t\left[\cos(ak_x)+2\cos\left(\frac{ak_x}{2}\right)
\cos\left(\frac{\sqrt{3}ak_y}{2}\right)\right].
\end{equation} 
The non-interacting local density of states (LDOS) 
$A_0(\omega)=\frac{1}{N}\sum_\mathbf{k}\delta(\omega-\varepsilon_\mathbf{k})$
can be in both cases calculated analytically
in terms of the complete elliptic integral
as explained in detail in Appendix~\ref{App:DOSzero}.

\subsection{Nambu Green function \label{SSec:NambuGF}}
From now we assume that the hybridization matrix elements are momentum-independent,
$V_{\mathbf{k}\sigma}=V_\sigma$ and drop the spin index 
$\sigma$ unless needed. 
We define the non-interacting ($U=0$) Green function as the resolvent of 
the non-interacting $\mathbf{k}$-resolved Hamiltonian 
that in the basis of $\psi_{\mathbf{k}}$ reads
\begin{equation}
\label{Eq:GF0}
G_0(\mathbf{k},z)=[zI-E_{\mathbf{k}}]^{-1}.
\end{equation}
Here $I$ is the $4\times 4$ unit matrix and $z$ is the complex energy.
Zeros of the determinant 
$\mathrm{Det}[G^{-1}_0(\mathbf{k},\omega+i0)]$ 
mark the poles of the Green function, i.e., the 
band structure of the non-interacting model as explained in detail
in Appendix~\ref{App:Nonint}.
The full interacting Green function is obtained from the Dyson equation
\begin{equation}
\label{Eq:DysonPAM}
G(\mathbf{k},z)=[G_0^{-1}(\mathbf{k},z)-\Sigma(\mathbf{k},z)]^{-1},
\end{equation}
where
\begin{equation}
\label{Eq:SelfEnergy}
\Sigma(\mathbf{k},z)=
\begin{pmatrix}
\Sigma_d(\mathbf{k},z) & 0 \\[0.3em]
0 & 0
\end{pmatrix}
\end{equation}
is the $4\times 4$ self-energy matrix. As the attractive interaction
in the conduction band is already incorporated on the static mean-field (BCS) level
into the non-interacting Green function, there is no explicit self-energy in that 
segment of the basis, although the conduction band is affected by 
$\Sigma_d$ via the hybridization $V$. 

\section{DMFT\label{Sec:DMFT}}
The DMFT maps the lattice model~\eqref{Eq:HamPAM} to an effective 
single-site dynamical model by performing a controlled limit to 
infinite spatial dimensions by proper scaling of the hopping 
parameters that guarantees the finiteness 
of the kinetic energy~\cite{Muller-Hartmann-1989}. In this limit, 
the correlation-induced self-energy becomes local~\cite{Metzner-1989}, 
$\Sigma_d(\mathbf{k},z)\rightarrow\Sigma_d(z)$. For finite lattice 
dimensions this is an approximation, $\Sigma_d(\mathbf{k},z)\approx\Sigma_d(z)$, 
however, it is the only approximation within 
the scheme. This approach, developed originally to solve the 
Hubbard model, can be successfully utilized
also for other lattice models including the 
SCPAM~\cite{Luitz-2010,Oei-2020}. 

As the DMFT equations for SCPAM were already derived in the above-mentioned 
works, we present here just a brief overlook of the procedure.
The effective single-site model in our case is the SCIAM,
\begin{equation}
\label{Eq:HamSCIAM}
\begin{aligned}
\mathcal{H}_{\mathrm{S}}&=
\sum_{\mathbf{k}\sigma}\tilde{\varepsilon}_{\mathbf{k}\sigma}
c_{\mathbf{k}\sigma}^\dag c_{\mathbf{k}\sigma}^{\phantom{\dag}}
-\tilde{\Delta}\sum_{\mathbf{k}}(c_{\mathbf{k}\uparrow}^\dag 
c_{\mathbf{\bar{k}}\downarrow}^\dag + \mathrm{H.c.})\\
&+\sum_{\sigma}\tilde{\varepsilon}_\sigma
d_{\sigma}^\dag d_{\sigma}^{\phantom{\dag}}
+Ud_{\uparrow}^\dag d_{\uparrow}^{\phantom{\dag}}
d_{\downarrow}^\dag d_{\downarrow}^{\phantom{\dag}}\\
&-\sum_{\mathbf{k}\sigma}(\tilde{V}_{\mathbf{k}\sigma}
c_{\mathbf{k}\sigma}^\dag d_{\sigma}^{\phantom{\dag}} + \mathrm{H.c.}).
\end{aligned}
\end{equation}
For simplicity, we use the same notation for the fermion operators 
($c$ for the conduction band and $d$ for the impurity) in the two models.
We can rewrite the SCIAM Hamiltonian in the Nambu formalism,
\begin{equation}
\begin{aligned}
\mathcal{H}_{\mathrm{S}}&=
D^\dag \tilde{E_d}D^{\phantom{\dag}}
+Ud_{\uparrow}^\dag d_{\uparrow}^{\phantom{\dag}}
d_{\downarrow}^\dag d_{\downarrow}^{\phantom{\dag}} \\
&+\sum_{\mathbf{k}}C_\mathbf{k}^\dag\tilde{E}_{c\mathbf{k}}C_\mathbf{k}^{\phantom{\dag}}
-\sum_{\mathbf{k}}(C_\mathbf{k}^\dag \tilde{V}_{\mathbf{k}}D^{\phantom{\dag}}+\mathrm{H.c}),
\end{aligned}
\end{equation}
where the spinors $D$ and $C_\mathbf{k}$ as well as the matrices 
$\tilde{E_d}$, $\tilde{E}_{c\mathbf{k}}$, and $\tilde{V}_{\mathbf{k}}$ 
are defined analogously to Eqs.~\eqref{Eq:NambuSpinors} and \eqref{Eq:NambuMatrices}.

We define the local element of the Matsubara (imaginary-frequency) 
Green function of the SCPAM,
\begin{equation}
\label{Eq:Gloc}
\begin{aligned}
G_{\mathrm{loc}}(i\omega_n)&=\frac{1}{N}\sum_{\mathbf{k}}G(\mathbf{k},i\omega_n)\\
&=
\begin{pmatrix}
G_{d,\mathrm{loc}}(i\omega_n) & G_{dc,\mathrm{loc}}(i\omega_n) \\[0.3em]
G^\dag_{dc,\mathrm{loc}}(i\omega_n) & G_{c,\mathrm{loc}}(i\omega_n)
\end{pmatrix},
\end{aligned}
\end{equation}
where $\omega_n=(2n+1)\pi k_BT$ is the $n$th fermionic Matsubara frequency and $T$ 
is the temperature. Following the standard DMFT procedure we define the bath Green 
function that serves as the input to the auxiliary problem 
by locally removing correlations from the local $d$-electron Green function,
\begin{equation}
\label{Eq:BathGF}
\mathcal{G}(i\omega_n)=[G_{d,\mathrm{loc}}^{-1}(i\omega_n)+\Sigma_d(i\omega_n)]^{-1}.
\end{equation}
Now we solve the auxiliary problem for the given value of $U$
with $\tilde{\Delta}$ determined by Eq.~\eqref{Eq:BCSdelta}
while $\tilde{\varepsilon}_\sigma$,
$\tilde{\varepsilon}_{\mathbf{k}\sigma}$
and $\tilde{V}_{\mathbf{k}\sigma}$ are encoded in the bath Green function $\mathcal{G}$.
As a result we obtain the local impurity Green function $G_{\mathrm{\mathrm{imp}}}(i\omega_n)$ 
and the impurity self-energy 
$\Sigma_{\mathrm{imp}}(i\omega_n)=\mathcal{G}^{-1}(i\omega_n)-G^{-1}_{\mathrm{imp}}(i\omega_n)$ 
and we identify it with the $d$-electron self-energy of SCPAM, 
$\Sigma_d(i\omega_n)=\Sigma_{\mathrm{imp}}(i\omega_n)$. 
Analogously to Eq.~\eqref{Eq:SelfEnergy} we define
\begin{equation}
\label{Eq:SelfEnergy2}
\Sigma(i\omega_n)=
\begin{pmatrix}
\Sigma_d(i\omega_n) & 0 \\[0.3em]
0 & 0
\end{pmatrix}.
\end{equation}
Using this self-energy and the Dyson equation 
\begin{equation}
\label{Eq:DysonPAMloc}
G(\mathbf{k},i\omega_n)=[G_0^{-1}(\mathbf{k},i\omega_n)-\Sigma(i\omega_n)]^{-1}
\end{equation}
we close the self-consistent loop. 
The convergence is achieved when $G_{d,\mathrm{loc}}(i\omega_n)=G_{\mathrm{imp}}(i\omega_n)$. 
The occupation numbers $n_d=n_{d\uparrow}+n_{d\downarrow}$, $n_{c}=n_{c\uparrow}+n_{c\downarrow}$
the induced pairing 
$\nu_d=\frac{1}{N}\sum_{\mathbf{k}}\langle d_{\mathbf{\bar{k}}\downarrow}d_{\mathbf{k}\uparrow}\rangle$ 
and the intrinsic pairing 
$\nu_c=\frac{1}{N}\sum_{\mathbf{k}}\langle c_{\mathbf{\bar{k}}\downarrow}c_{\mathbf{k}\uparrow}\rangle$ 
can be then calculated from the local Green function, Eq.~\eqref{Eq:Gloc},
\begin{equation}
\label{Eq:densities}
\frac{1}{\beta}\sum_n G_{\mathrm{loc}}(i\omega_n)=
\begin{pmatrix}
n_{d\uparrow} & \nu_{d}           & n_{dc}        & \nu_{dc} \\
\nu_{d}       & 1-n_{d\downarrow} & -\nu_{dc}     & -n_{dc} \\
n_{dc}        & -\nu_{dc}         & n_{c\uparrow} & \nu_{c} \\
\nu_{dc}      & -n_{dc}           & \nu_{c}       & 1-n_{c\downarrow}
\end{pmatrix}.
\end{equation}

The auxiliary single-site problem can be solved using either 
numerically exact but expensive techniques like QMC, numerical renormalization group (NRG)
or the exact diagonalization, or
in an approximative way using simpler but faster solvers based on diagrammatic 
expansion techniques like IPT, non-crossing approximation or various 
slave-boson techniques~\cite{Kotliar-2006}. We chose the CT-HYB QMC method
to calculate the overall properties of SCPAM, backed up with approximate 
spectral functions provided by the IPT.

\subsection{CT-HYB\label{SSec:CTHYB}}
We use the CT-HYB QMC technique~\cite{Gull-2011} as a numerically exact solver for the SCIAM. 
As the Hamiltonian in Eq.~\eqref{Eq:HamSCIAM} does not conserve 
particle number, we use a standard trick where we perform 
a canonical particle-hole transformation in the spin-down segment of 
the Hilbert space~\cite{Luitz-2010},
$(d_{\uparrow}^\dag,d_{\uparrow}^{\phantom{\dag}},
d_{\downarrow}^\dag,d_{\downarrow}^{\phantom{\dag}})\rightarrow
(d_{\uparrow}^\dag,d_{\uparrow}^{\phantom{\dag}},
d_{\downarrow}^{\phantom{\dag}},d_{\downarrow}^\dag)$
and 
$(c_{\mathbf{k}\uparrow}^\dag,c_{\mathbf{k}\uparrow}^{\phantom{\dag}},
c_{\mathbf{k}\downarrow}^\dag,c_{\mathbf{k}\downarrow}^{\phantom{\dag}})\rightarrow
(c_{\mathbf{k}\uparrow}^\dag,c_{\mathbf{k}\uparrow}^{\phantom{\dag}},
c_{\mathbf{\bar{k}}\downarrow}^{\phantom{\dag}},c_{\mathbf{\bar{k}}\downarrow}^\dag)$. 
This transformation maps SCIAM to SIAM with attractive interaction $-U$
and changes the sign of the energy levels of the spin-down electrons,
$\tilde{\varepsilon}_\sigma\rightarrow\sigma\tilde{\varepsilon}_\sigma$
and $\tilde{\varepsilon}_{\mathbf{k}\sigma}
\rightarrow\sigma\tilde{\varepsilon}_{\mathbf{k}\sigma}$.
The resulting Hamiltonian is conserving and can be treated using standard 
solvers.

CT-HYB is an inherently finite-temperature method that measures the Green 
function in imaginary-time domain $G(\tau)$. Therefore, the spectral functions are 
not accessible without performing an analytic continuation to real frequencies 
which, for stochastic data, is a notoriously
ill-defined problem~\cite{Jarrell-1996}. However, the value of the spectral 
function at the Fermi energy $A(\omega=0)$ can be approximated at low temperatures 
by $\beta G(\tau=\beta/2)$~\cite{Liebsch-2003}
where $\beta=1/k_BT$ is the inverse temperature. Since
\begin{equation}
G(\tau)=\int_{-\infty}^\infty \!\!d\omega
\frac{e^{-\tau\omega}}{1+e^{-\beta\omega}}A(\omega),
\end{equation}
we get
\begin{equation}
\label{Eq:GBeta2}
G(\beta/2)=\int_{-\infty}^\infty \!\!d\omega
\frac{A(\omega)}{2\cosh(\beta\omega/2)},
\end{equation}
i.e., that $G(\beta/2)$ is a measure of the integrated spectral 
weight on an interval of few $k_BT$ around the Fermi energy. As 
$(\beta/2)\cosh^{-1}(\beta\omega/2)\rightarrow
\pi\delta(\omega)$ for $\beta\rightarrow\infty$,
we arrive to a simple expression $\beta G(\beta/2)\approx \pi A(0)$ 
valid for very low temperatures. This measure is often used to 
locate metal-insulator transitions in Hubbard-like models and we can utilize
it to identify the possible crossing of the impurity bands at the Fermi 
energy.

\subsection{IPT\label{SSec:IPT}}
The drawback of the CT-HYB method coming from the inability to provide 
spectral functions with adequate resolution hinders its usability to 
describe experiments performed using scanning tunneling spectroscopy techniques.
Therefore we used the IPT to provide an approximate 
shape of the spectral function.

\begin{figure}[ht]
\includegraphics[width=0.92\columnwidth]{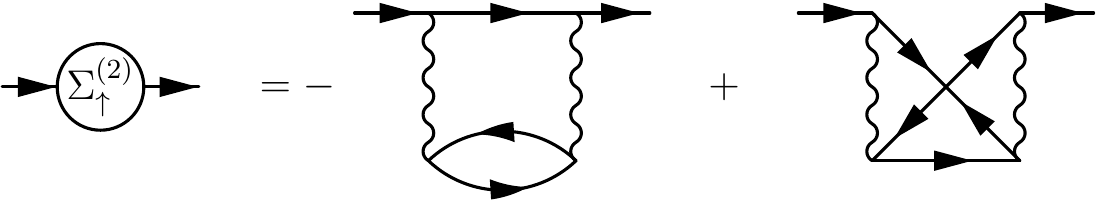}\\
\includegraphics[width=0.92\columnwidth]{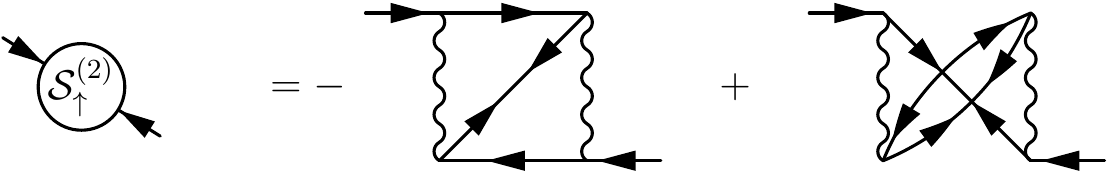}
\caption{Diagrammatic representation of the two spin-up components of the 
second-order correction to the $d$-electron self-energy, Eq.~\eqref{Eq:Sigma2ndPT}.
Top line represents spin-up, bottom line spin-down. Arrows with single heads 
(diagonal elements of the bath Green function) mark the propagation of electrons 
(right arrow) and holes (left arrow), arrows with double heads (off-diagonal
elements of the bath Green function) represent creation and annihilation 
of the Cooper pairs. Vertical wavy line is the Coulomb interaction vertex $U$.
\label{Fig:SEdiag}}
\end{figure}

IPT uses the second-order perturbation theory (2PT) in the interaction strength $U$
as the solver for the impurity problem and 
was originally introduced to solve the Hubbard model at half-filling~\cite{Georges-1992}. 
It was successfully used to solve the 
PAM~\cite{Schweitzer-1991,Rozenberg-1996,Vidhyadhiraja-2000} as well 
as the Hubbard model with BCS superconducting bath~\cite{Garg-2005,Koley-2017}.
The dynamical self-energy reads
\begin{equation}
\label{Eq:SigmaIPT}
\Sigma_{\mathrm{imp}}(i\omega_n)=\Sigma^{\mathrm{HF}}+\Sigma^{(2)}(i\omega_n),
\end{equation}
where $\Sigma^{\mathrm{HF}}$ is the static Hartree-Fock self-energy 
\begin{equation}
\label{Eq:SigmaHF}
\Sigma^{\mathrm{HF}}=\frac{U}{\beta}\sum_n
\begin{pmatrix}
\mathcal{G}_{\downarrow}(i\omega_n) & \mathcal{F}_{\downarrow}(i\omega_n) \\[0.3em]
\bar{\mathcal{F}}_{\uparrow}(i\omega_n) & \bar{\mathcal{G}}_{\uparrow}(i\omega_n)
\end{pmatrix}
\end{equation} 
and $\Sigma^{(2)}(i\omega_n)$ is the second-order correction that can
be easily written down in the imaginary-time domain and reads~\cite{Garg-2005}
\begin{equation}
\label{Eq:Sigma2ndPT}
\begin{aligned}
\Sigma^{(2)}_\sigma(\tau)&=-U^2\bar{\mathcal{G}}_{-\sigma}(-\tau)
[\mathcal{G}_\uparrow(\tau)\bar{\mathcal{G}}_\downarrow(\tau)
-\mathcal{F}_\uparrow(\tau)\bar{\mathcal{F}}_\downarrow(\tau)], \\
\mathcal{S}^{(2)}_\sigma(\tau)&=-U^2\mathcal{F}_{-\sigma}(-\tau)
[\mathcal{G}_\uparrow(\tau)\bar{\mathcal{G}}_\downarrow(\tau)
-\mathcal{F}_\uparrow(\tau)\bar{\mathcal{F}}_\downarrow(\tau)], \\
\end{aligned}
\end{equation}
where we denoted the elements of the local $d$-electron 
bath Green function~\eqref{Eq:BathGF} and self-energy as
\begin{equation}
\label{Eq:GDloc}
\mathcal{G}=
\begin{pmatrix}
\mathcal{G}_{\uparrow} & \mathcal{F}_{\uparrow} \\[0.3em]
\bar{\mathcal{F}}_{\downarrow} & \bar{\mathcal{G}}_{\downarrow}
\end{pmatrix},\quad
\Sigma^{(2)}=
\begin{pmatrix}
\Sigma^{(2)}_{\uparrow} & \mathcal{S}^{(2)}_{\uparrow} \\[0.3em]
\bar{\mathcal{S}}^{(2)}_{\downarrow} & \bar{\Sigma}^{(2)}_{\downarrow}
\end{pmatrix}
\end{equation}
and bar denotes a charge-conjugate (hole) function. 
The relation between functions in imaginary frequency and imaginary time reads
\begin{equation}
X(\tau)=\frac{1}{\beta}\sum_n e^{-i\omega_n\tau}X(i\omega_n)
\end{equation}
and we use the same notation for both.
Diagrammatic representation of the dynamic part of the self-energy 
is depicted in Fig.~\ref{Fig:SEdiag}.
The elements with inverted spin can be obtained via symmetry relations 
$\mathcal{G}_{\sigma}(i\omega_n)=-\bar{\mathcal{G}}_{-\sigma}(-i\omega_n)$
and 
$\mathcal{F}_{\sigma}(i\omega_n)=\bar{\mathcal{F}}_{-\sigma}(i\omega_n)$. 

The IPT was later generalized for arbitrary filling~\cite{Kajueter-1995}
by introducing interpolation parameters $\mathcal{A}$ and $\mathcal{B}$,
\begin{equation}
\label{Eq:SigmaIPT_AB}
\Sigma_{\mathrm{imp}}(i\omega_n)=\Sigma^{\mathrm{HF}}+\mathcal{A}\Sigma^{(2)}(i\omega_n)
\left[1-\mathcal{B}\Sigma^{(2)}(i\omega_n)\right]^{-1}.
\end{equation}
Parameter $\mathcal{A}$ is obtained from the exact asymptotics in the
high-frequency limit ($\omega_n\rightarrow\pm\infty$) of the spectral function
that can be calculated from the equation of motion~\cite{Geipel-1988,Nolting-1989}
and $\mathcal{B}$ is derived from the atomic limit $t\rightarrow 0$.
The matrix form of the parameters for superconducting models in the 
Nambu formalism was derived in Ref.~\cite{Garg-2005}
where the authors showed that matrix $\mathcal{B}$ is zero in the superconducting case.

Our experience with 2PT calculations of SCIAM~\cite{Zonda-2015,Zonda-2016}
shows that a more viable way how to modify the behavior of the self-energy is the
correction of the static (Hartree-Fock) part $\Sigma^{\mathrm{HF}}$, so that the initial 
occupation matrix calculated from the bath Green function matches the final one
calculated from the 2PT propagator. In this method, the dynamical self-energy~\eqref{Eq:Sigma2ndPT}
is calculated only once from the bath Green function $\mathcal{G}$, but the
static part $\Sigma^{\mathrm{HF}}$ is consistently recalculated from Eq.~\eqref{Eq:SigmaHF}
in which the bath propagator $\mathcal{G}$ is replaced by the 2PT propagator
\begin{equation}
\label{Eq:2PT_GF}
G^{(2)}_{d,\mathrm{loc}}(i\omega_n)=[\mathcal{G}^{-1}(i\omega_n)
-\Sigma^{\mathrm{HF}}-\Sigma^{(2)}(i\omega_n)]^{-1}.
\end{equation}
In theory, a fully self-consistent update is possible, where
the $\Sigma^{(2)}$ is also calculated iteratively from $G^{(2)}_{d,\mathrm{loc}}$. 
Our experience, however, is that this method is numerically 
much more demanding and can lead to spurious behavior if used inside
of a DMFT loop.

We implemented the IPT method in the Matsubara frequency formalism.
While it is possible to implement the IPT solver directly in real frequencies, 
the complicated sub-gap structure of the impurity Green function makes such calculations 
problematic as one has to carefully identify the positions of the in-gap states
while the local spectral functions $A_d(\omega)=-\Imm G_{d,\mathrm{loc}}(\omega+i0)/\pi$
and $A_c(\omega)=-\Imm G_{c,\mathrm{loc}}(\omega+i0)/\pi$
can be obtained reliably from the Matsubara Green function using the Pad\'e analytic 
continuation technique~\cite{Vidberg-1977}.

Properties of 2PT solution for SCIAM were studied in Refs.~\cite{Zonda-2015,Zonda-2016}, 
showing that 2PT with corrected static self-energy as described above 
provides reliable results compared to NRG and QMC for 
weak and intermediate coupling $U$ if the ground state of the impurity 
model is a singlet, but it fails for the doublet ground state. 
As the ground state for the non-interacting ($U=0$) model is always a singlet, 
there is no way to switch on adiabatically the interaction and end up in the doublet state
as it is separated from the singlet state by a quantum critical point. 
Also, it is not possible to perform a diagrammatic expansion around a doublet 
(or any multiplet) ground state without prior lifting of the degeneracy, 
e.g., by magnetic field. Therefore 2PT gives a non-physical solution for 
SCIAM in the doublet phase with the in-gap states pinned at the Fermi energy.
This is indeed a serious limitation and one 
has to keep an eye on the sign of the induced gap $\nu_d$. 
Negative values of this parameter mark the situations where IPT becomes
unreliable due to the above-mentioned failure of the underlying 
2PT impurity solver, even though the results can show reasonable
agreement with the numerically exact QMC data.

\section{Results \label{Sec:Results}}
Calculations were performed using our own DMFT code based on
the TRIQS 2.2 libraries~\cite{Parcollet-2015}
and the TRIQS/CTHYB hybridization-expansion solver~\cite{Seth-2016} 
at temperature $k_BT=0.025t$ with
a cutoff in Matsubara frequencies $\omega_n^{\mathrm{max}}\geq 300t$.
Few data sets were recalculated at lower temperature to assess
its effect on the position of the phase boundaries.
We restrict our results to the half-filled situation $n_c+n_d=2$.
The DMFT cycle was started
from zero self-energy and a high value of the gap parameter $\Delta$
which was recalculated in each DMFT iteration
\footnote{
It is worth noting that our method cannot identify a possible superconducting state 
that forms without explicit attractive pairing, i.e., at $g=0$ as discussed by 
Bodensiek \textit{et al.} in Ref.~\cite{Bodensiek-2013}. Our order parameter 
$\Delta=g\nu_c$ is recalculated in each DMFT iteration from the intrinsic pairing
$\nu_c$ and therefore it is always proportional to $g$. As a result, we cannot rule
out the existence of another SC phase where the pairing would be mediated by
magnetic excitations as hypothesized in the above-mentioned paper.}
together with the 
chemical potential $\mu$ that fixes the total filling.
The CT-HYB solver was used to obtain numerically exact results on the
impurity density matrix, from which the occupancy, double occupancy and the 
induced pairing was calculated. These results were recalculated using IPT,
showing good agreement, at least for the square lattice. The spectral function 
$A(\omega)$ was then obtained from the IPT Green function using the Pad\'e 
analytic continuation method. An imaginary frequency offset 
$\eta=5.10^{-3}$ was used to guarantee the correct analytic behavior 
of the spectral functions.

\subsection{Phase diagrams\label{SSec:Phasediag}}

\begin{figure}[ht]
\includegraphics[width=\columnwidth]{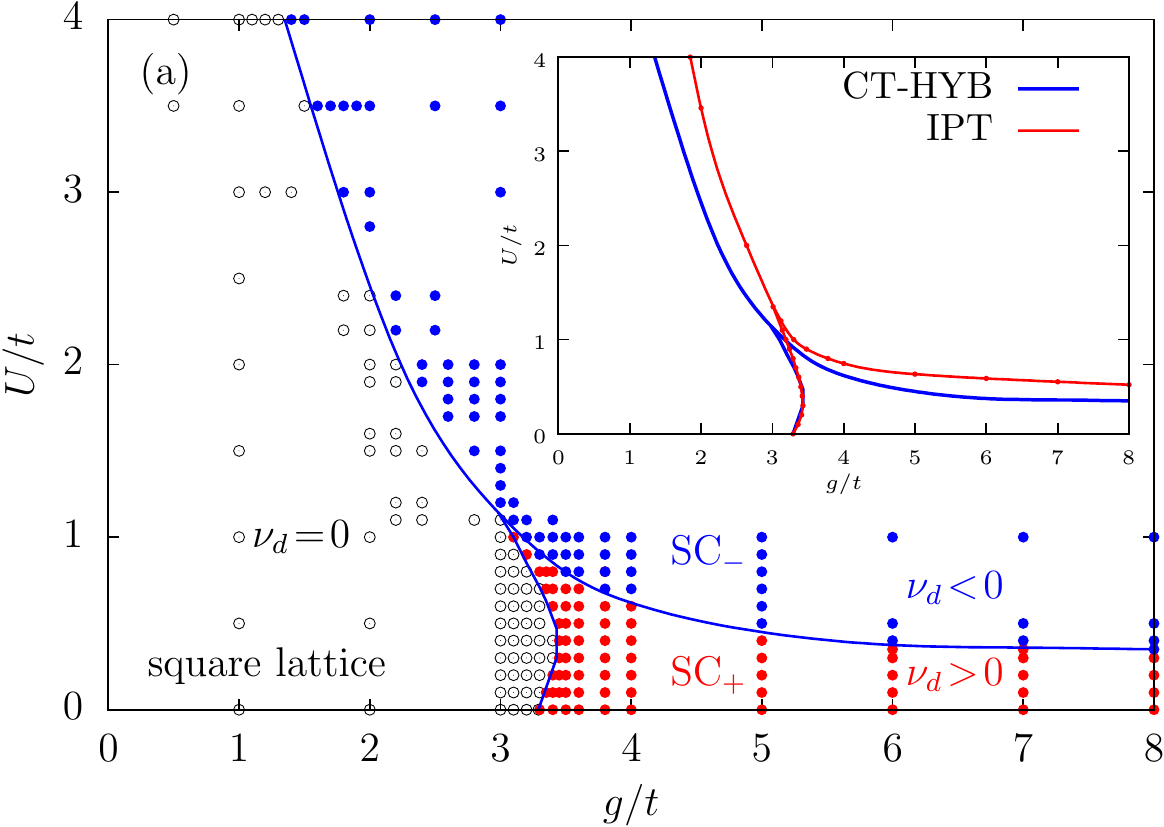}
\includegraphics[width=\columnwidth]{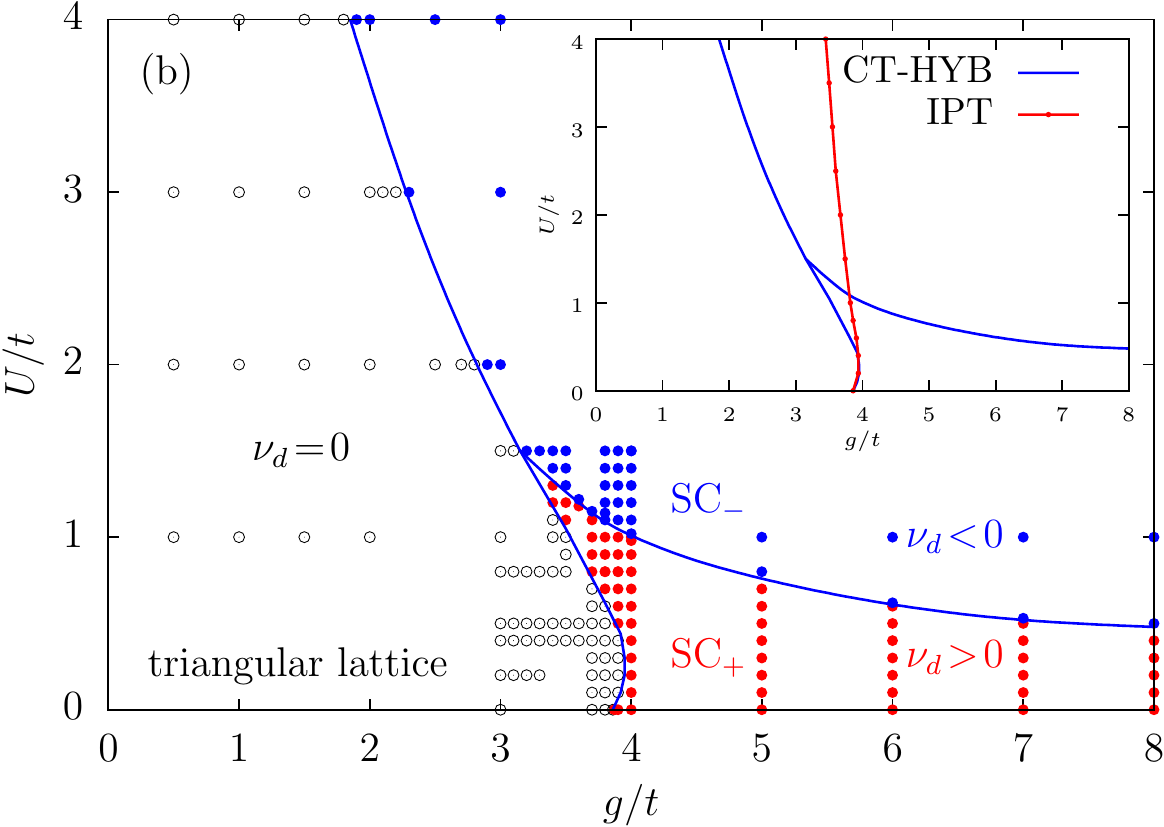}
\caption{Phase diagram of SCPAM in the $g-U$ plane for $V=0.5t$ and $k_BT=0.025t$
on (a) square lattice and (b) triangular lattice
calculated using CT-HYB as the DMFT solver. 
Bullets represent individual DMFT calculations. 
Black: Non-superconducting, Kondo insulator phase.
Red (blue): Superconducting phase with positive (negative) induced pairing.
Blue lines are approximate position of the phase boundaries.
Inset: Comparison of the position of the phase boundary calculated
using CT-HYB (blue) and IPT (red).
\label{Fig:PhaseDiag}}
\end{figure}

In Fig.~\ref{Fig:PhaseDiag} we plot the phase diagram of SCPAM in the $g-U$ 
plane on a square, Fig.~\ref{Fig:PhaseDiag}(a),
and triangular, Fig.~\ref{Fig:PhaseDiag}(b), lattice at half-filling for $V=0.5t$. 
We choose this value of the hybridization $V$ as the superconducting
correlations are strongest here in the non-interacting ($U=0$) case
(see Appendix~\ref{App:Nonint2} for mode details).
The two phase diagrams look rather similar, largely due to the local 
nature of the DMFT. The main panel shows the phase diagram calculated 
using CT-HYB as the DMFT solver. Each individual bullet represents 
a separate calculation. The black empty bullets mark the 
non-superconducting, KI phase. The superconducting region can be separated into two 
phases that we mark $\mathrm{SC_+}$ (red bullets) and $\mathrm{SC_-}$ (blue bullets)
by the sign of the induced paring $\nu_d$. Solid lines represent the 
approximate position of the phase boundaries. 

\begin{figure}[ht]
\includegraphics[width=\columnwidth]{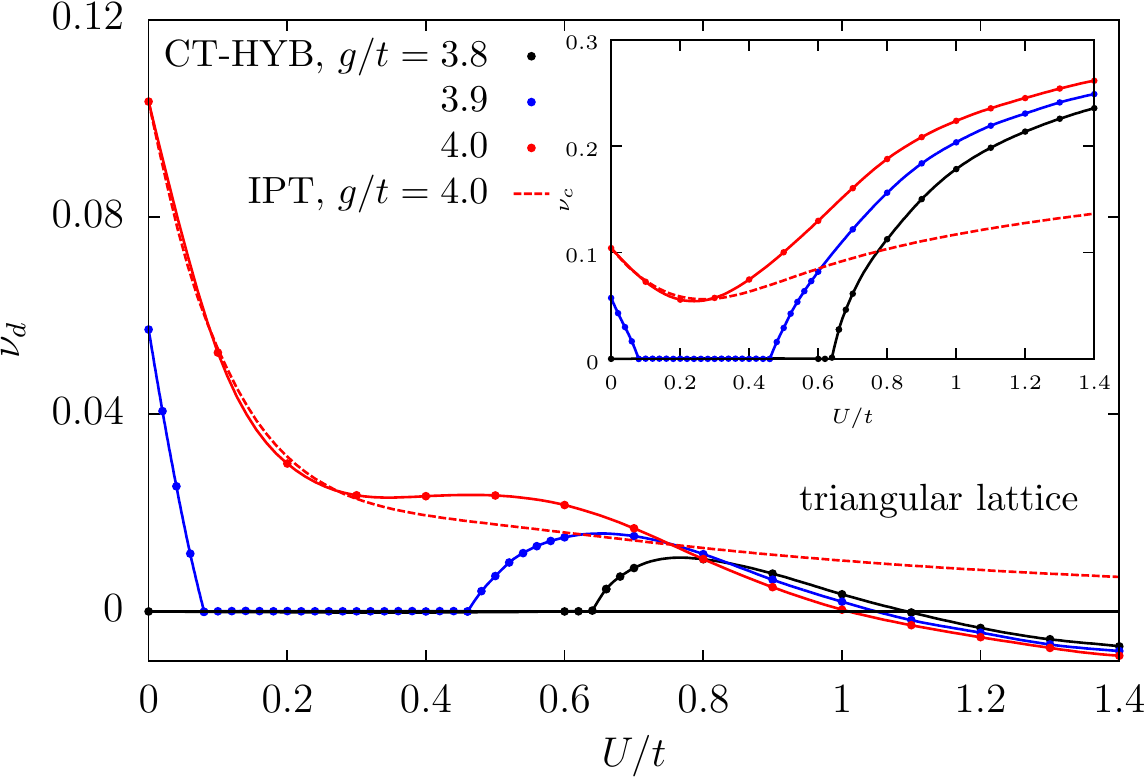}
\caption{Induced gap $\nu_d$ (main panel) and the intrinsic pairing $\nu_c$
(inset) as functions of the interaction strength $U$
calculated using CT-HYB as the DMFT solver on the triangular lattice
for three values of the electron-phonon coupling $g$ 
in the vicinity of the Kondo insulator - superconductor phase boundary. 
The system shows a reentrant behavior for $g=3.9t$ (blue). Bullets
represent quantum Monte Carlo data, solid lines are splines 
and serve only as a guide for the eye. The dashed line represents the IPT 
results.
\label{Fig:gdep_tri_cthyb}}
\end{figure}

In the non-interacting ($U=0$) case the system is a KI
for $g<g_{c0}$ and a  $\mathrm{SC_+}$ superconductor for $g>g_{c0}$.
The critical value of the attractive coupling is 
$g_{c0}\approx 3.30t$ for the square and $g_{c0}\approx 3.86t$ for 
the triangular lattice
~\footnote{The existence of the critical value of $g$ at $U=0$ is connected 
with the fact that at half-filling the normal ($g=0$) state is an 
insulator. For fillings where the normal state is metallic at $U=0$ 
the system is a superconductor for any $g>0$, assuming the 
temperature is smaller than the critical temperature.}. 
By increasing the interaction strength $U$,
the critical coupling $g_c$ shows a reentrant behavior as described later
in Fig.~\ref{Fig:gdep_tri_cthyb}. The transition between the KI and $\mathrm{SC_+}$ phases 
at constant $U$ is continuous and BCS-like, i.e., $\Delta\approx(g-g_c)^{1/2}$.
The character of the transition changes at the point where the KI-superconductor
transition line meets the transition line that separates the $\mathrm{SC_+}$ and 
$\mathrm{SC_-}$ phases. The transition from KI to $\mathrm{SC_-}$ phase 
is discontinuous with a jump in the order parameter $\Delta$. 
The $\mathrm{SC_+}$ and $\mathrm{SC_-}$ phases are separated by a smooth crossover
that becomes sharper with decreasing temperature and the line in the phase diagram 
that separates these phases marks the point where the induced gap $\nu_d$ changes sign.
The behavior of this crossover with decreasing temperature and a possible method 
how to extrapolate the transition to zero temperatures 
is discussed in Sec.~\ref{SSec:ExpOrder}. 

All the presented DMFT calculations use zero self-energy and a large value of 
$\Delta$ as the initial condition.
In the search for the expected hysteresis behavior as described by Luitz 
and Assaad in Ref.~\cite{Luitz-2010} we performed 
a second calculation for selected values of the coupling strength $g$,
starting from the self-energy for large interaction strength $U>4t$.
We encountered no measurable difference within the QMC error bars 
between the results at $k_BT=0.025t$. This is consistent with the conclusions 
of Ref.~\cite{Luitz-2010} where the authors had to perform the calculation at very 
low temperatures $k_BT\approx0.007t$ to be able to observe any measurable 
hysteresis.

The insets show the phase boundaries calculated using 
IPT solver (red) compared to the CT-HYB result from the main 
panels (blue). For the square lattice IPT overestimates the 
value of the critical interaction strength $U$ but provides qualitatively
correct topology of the phase diagram. For the triangular 
lattice IPT again gives a fair guess for the 
position of the KI-superconductor phase boundary, but fails to 
predict the change of the sign of the induced gap which is positive
for all parameters and only decreases towards zero with increasing 
interaction strength $U$. 
This unsatisfactory behavior of the IPT for triangular lattice is 
connected with the failure of the solver to provide reasonable 
data away from half filling. As the triangular lattice is not bipartite, 
the fixed total filling $n_c+n_d=2$ does not guarantee that the individual 
bands are half-filled and the filling of the $d$ band $n_d$
changes with the model parameters due to the change of the chemical 
potential $\mu$. The IPT overestimates the deviation from half-filling of 
the $d$-band. Therefore the correlation effects caused by increasing 
$U$, which are strongest at half-filling, are damped, keeping
the the system in the $\mathrm{SC_+}$ phase for all values of $U$.

\begin{figure*}[ht]
\includegraphics[width=2\columnwidth]{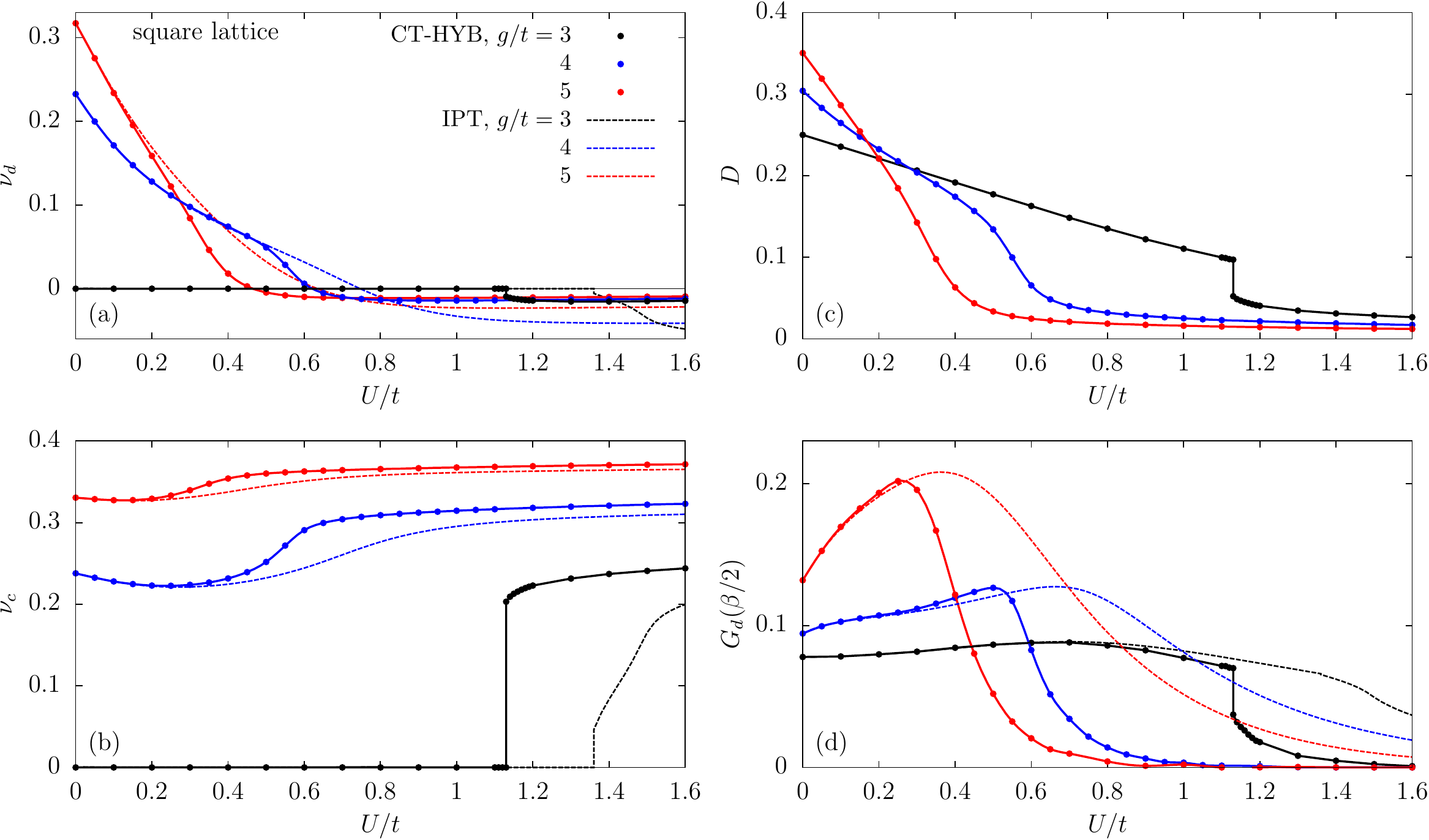}
\caption{Induced pairing $\nu_d$ (a), intrinsic pairing $\nu_c$ (b), 
$d$-band double occupancy $D$ (c) and the value of the diagonal element 
of the imaginary-time Green function $G_{d,loc}(\tau)$ at $\tau=\beta/2$ (d)
as functions of interaction strength $U$ for three values of the coupling
$g$ calculated using CT-HYB (bullets) and IPT (dashed lines) 
as the DMFT solvers for SCPAM on the square lattice for $V=0.5t$ 
and $k_BT=0.025t$. Solid lines are splines of the quantum Monte Carlo data 
and serve only as a guide for the eye.
\label{Fig:Udep_sq}}
\end{figure*}

In Fig.~\ref{Fig:Udep_sq} we plot the comparison of CT-HYB and IPT results
along three cuts of the phase diagram for the square lattice, 
Fig.~\ref{Fig:PhaseDiag}(a), for $g=3t$ (black), $4t$ (blue) and $5t$ (red).
The bullets represent CT-HYB results, dashed lines IPT results. Figs.~\ref{Fig:Udep_sq}(a)
and \ref{Fig:Udep_sq}(b) show the behavior of the induced pairing $\nu_d$ 
and the intrinsic pairing $\nu_c$
\footnote{We plot the pairing instead of the gap because the data for
pairing are more commensurate for different values of $g$ and fit better
in a single plot.}. 
For $g=3t$ we start at $U=0$ from the KI phase with zero pairing.
The transition to the $\mathrm{SC_-}$ phase is discontinuous with a jump in the
order parameter at $U_c\approx 1.13t$. IPT overestimates this value 
by roughly $20\%$. The different height of the jump is also due to the
proximity to the 'triple point' where the two transition lines meet.
For $g=4t$ and $5t$ the induced pairing decreases 
with increasing interaction strength, changing sign from positive 
to negative. The intrinsic $\nu_c$ shows a slight kink around that point,
but otherwise it is largely independent of the interaction strength.
IPT again overestimates the position of the transition point but provides a 
reasonable fit to the CT-HYB data in both superconducting phases, 
despite the fact it becomes unreliable in situations where $\nu_d$ is negative.

Fig.~\ref{Fig:Udep_sq}(c) shows the double occupancy of the $d$-band 
$D=\langle d^\dag_\uparrow d^{\phantom{\dag}}_\uparrow
d^\dag_\downarrow d^{\phantom{\dag}}_\downarrow\rangle$
calculated using CT-HYB.
Its value at $U=0$ is 1/4 for $g=3t$ and larger than 1/4 in the superconducting
phase due to the attractive interaction induced in the impurity band by the 
proximity effect. It decreases with the increasing interaction strength $U$ as 
the doubly occupied state becomes energetically more expensive and shows a sharp 
downturn at the crossover to the $\mathrm{SC_-}$ phase, where the ground 
state of the impurity model is a doublet of singly-occupied states.

Fig.~\ref{Fig:Udep_sq}(d) shows value of the $d$-electron imaginary-time Green 
function $G_d(\tau)$ at $\tau=\beta/2$, Eq.~\eqref{Eq:GBeta2}, which is a 
measure of the spectral weight in the narrow window of a few $k_BT$ around the Fermi energy.
For $g=3t$ the system is a KI with a narrow gap smaller than the
relevant energy window so this value is small but non-zero, 
decreasing sharply at the transition point to the superconductor
where the additional gap of width $\Delta$ opens at the Fermi energy,
pushing the spectral weight to higher energies.
As expected, IPT fits this value very well in the KI phase.
For $g=4t$ and $5t$ this function shows a peak before 
the transition point, then decreases rapidly,
suggesting that the subgap Andreev bands are approaching the
Fermi energy at the crossover in a similar manner
as the ABS are crossing in the SCIAM at the $0-\pi$ (singlet-doublet)
transition. This feature is discussed in more detail in the next section.

To illustrate the reentrant behavior of the superconductivity 
and the failure of IPT for the triangular lattice we plotted
in Fig.~\ref{Fig:gdep_tri_cthyb} the induced pairing $\nu_d$ (main panel) 
and the intrinsic pairing $\nu_c$ (inset) as functions of the 
interaction strength $U$ for triangular lattice calculated using CT-HYB 
(bullets) for three values of $g$ close to the KI-$\mathrm{SC_+}$ phase boundary
and added the IPT result for $g=4t$ (dashed line). For $g=3.9t$ the 
superconducting order is quickly suppressed by the increasing Coulomb interaction
just to re-emerge again at higher values of $U$. Similar reentrant features are
discussed in Ref.~\cite{Oei-2020}. The $d$-band occupation is $n_d\approx 1.15$ at 
$U=0$ for all three values of $g$ and according to the CT-HYB result it quickly approaches 
unity (half filling) with increasing interaction strength.
This enhances the correlation effects and eventually drives the system into the 
$\mathrm{SC_-}$ phase.
On the other hand, IPT result shows only very slow decrease of the $d$ band 
occupation. As a result, it fails to predict the change of the sign of $\nu_d$ which only 
asymptotically approaches zero with increasing interaction strength.
A more elaborate modification of the IPT algorithm is needed to correctly
describe the change of the occupation to study the model on non-bipartite lattices,
although this cannot overcome the principal problem of the method which is the inability
to describe the spinful ground state of the impurity problem as discussed 
in Sec.~\ref{SSec:IPT} which seriously limits the reliability of this
method for certain superconducting models.

\subsection{IPT spectral functions\label{SSec:Spectra}}

\begin{figure}[ht]
\includegraphics[width=\columnwidth]{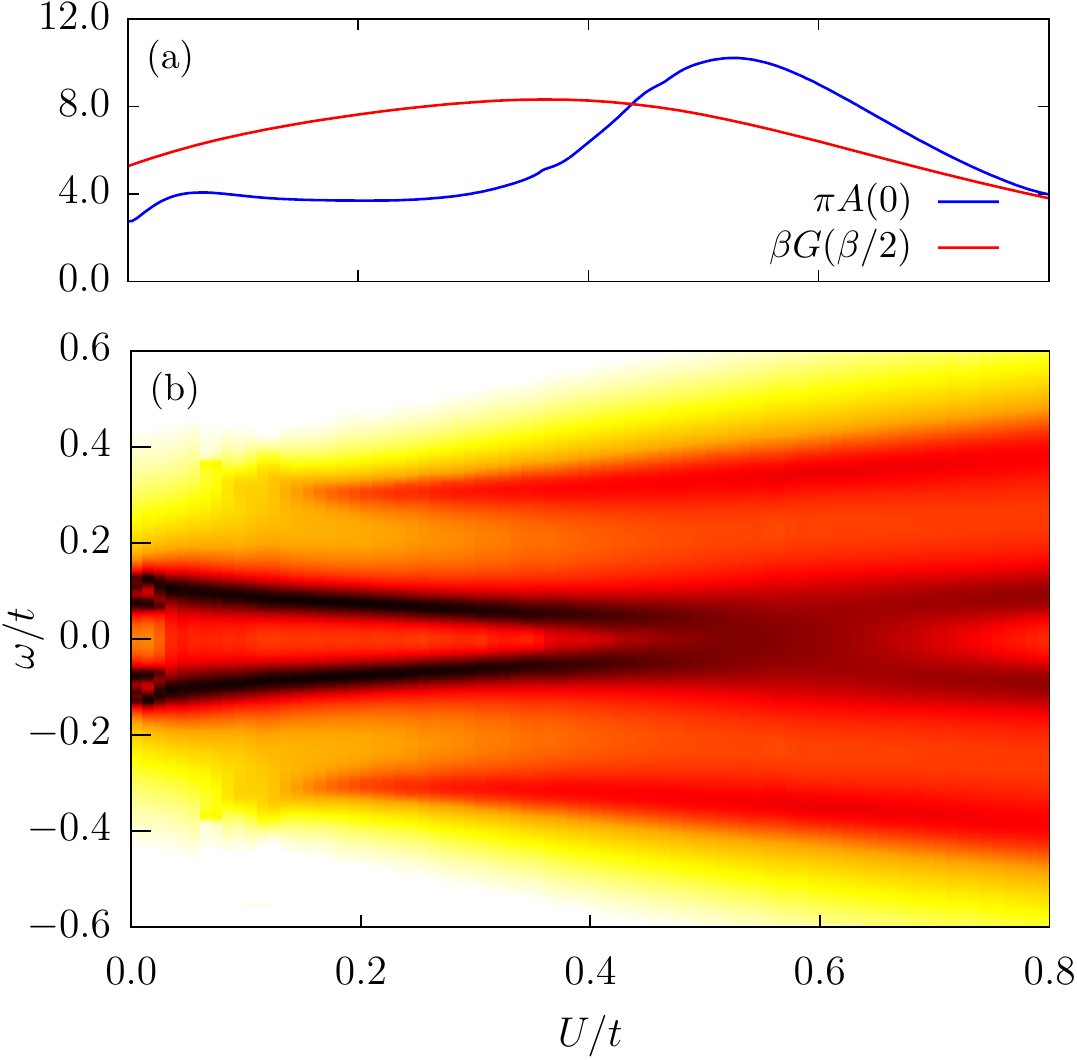}
\caption{(a) Value of the diagonal element of the total spectral function 
$A(\omega)$ at the Fermi energy $\omega=0$ (blue) as a function of the interaction 
strength $U$ calculated using IPT for $V=0.5t$, $g=5t$, and $k_BT=0.025t$
compared to the $\beta G(\beta/2)$ from Fig.~\ref{Fig:Udep_sq}(d) (red). 
(b) Heatmap of  $A(\omega)$ for the same parameters as in panel (a).
\label{Fig:spec_map_g5}}
\end{figure}

\begin{figure}[ht]
\includegraphics[width=\columnwidth]{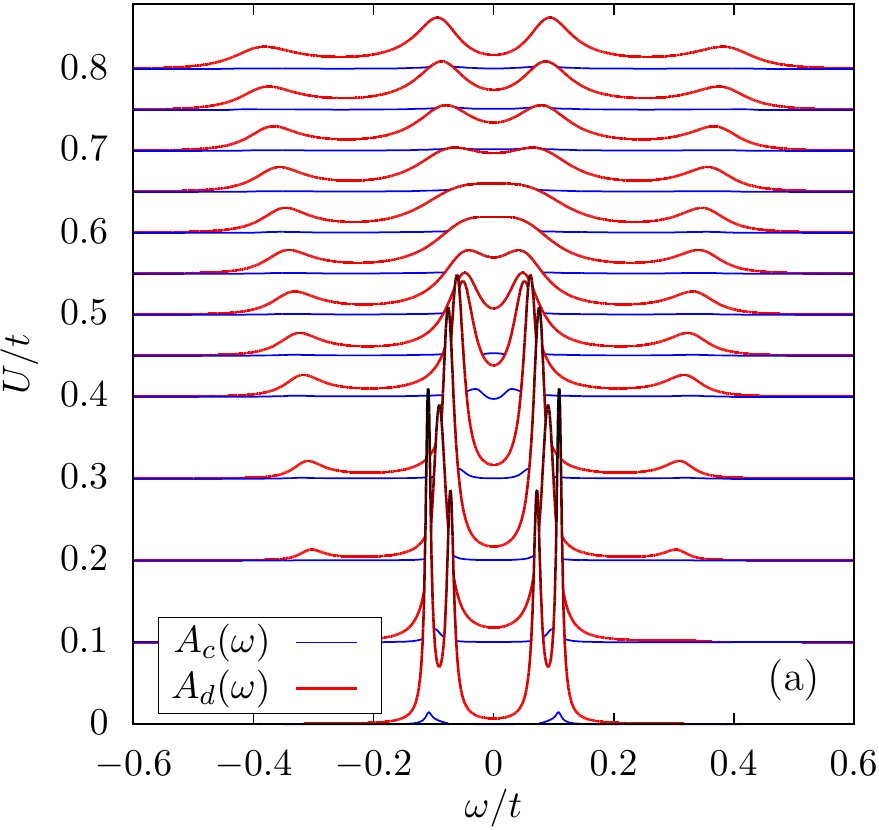}
\caption{In-gap spectral functions $A_c(\omega)$ (blue) 
and $A_d(\omega)$ (red) calculated using the IPT solver
for selected values of interaction strength $U$ for the same parameters
as in Fig.~\ref{Fig:spec_map_g5}, $V=0.5t$, $g=5t$, and $k_BT=0.025t$.
\label{Fig:spec_cuts_g5}}
\end{figure}

The transition between the two superconducting phases can be further illustrated on
the behavior of the in-gap bands in the spectral function. Here we limit our
results to the square lattice where IPT provides reasonable agreement with CT-HYB.
In Fig.~\ref{Fig:spec_map_g5} we plot the  
diagonal part of the total spectral function $A(\omega)=A_d(\omega)+A_c(\omega)$
calculated using IPT
as a function of the interaction strength $U$ for $V=0.5t$, $g=5t$, and $k_BT=0.025t$
at the Fermi energy $\omega=0$ (a) together with the heatmap of $A(\omega)$
in the region around the Fermi energy (b). We also plot in 
Fig.~\ref{Fig:spec_cuts_g5} the spectral function for selected values of $U$
to better specify the shape of the in-gap bands. The gap edges lie 
at $\pm\Delta=\pm g\nu_c\sim\pm1.7t$ which follow the red 
dashed line from Fig.~\ref{Fig:Udep_sq}b (scaled by $g=5t$). 
As expected, the in-gap spectral function
is largely dominated by the $d$-band contribution. The double-peak structure of 
the individual bands is quickly smeared out by the interaction strength
and the two bands move closer together, losing coherence,
merging at $U\approx 0.55t$ and move apart again. The merging of 
the bands roughly coincides with the zero of the induced pairing
$\nu_d$ (red dashed line in Fig.~\ref{Fig:Udep_sq}(a)) at $U\approx0.64t$.
Furthermore, two additional bands are formed at higher energies
that resemble Hubbard bands of the SIAM as they move to larger energies 
with increasing $U$.

\begin{figure}[ht]
\includegraphics[width=\columnwidth]{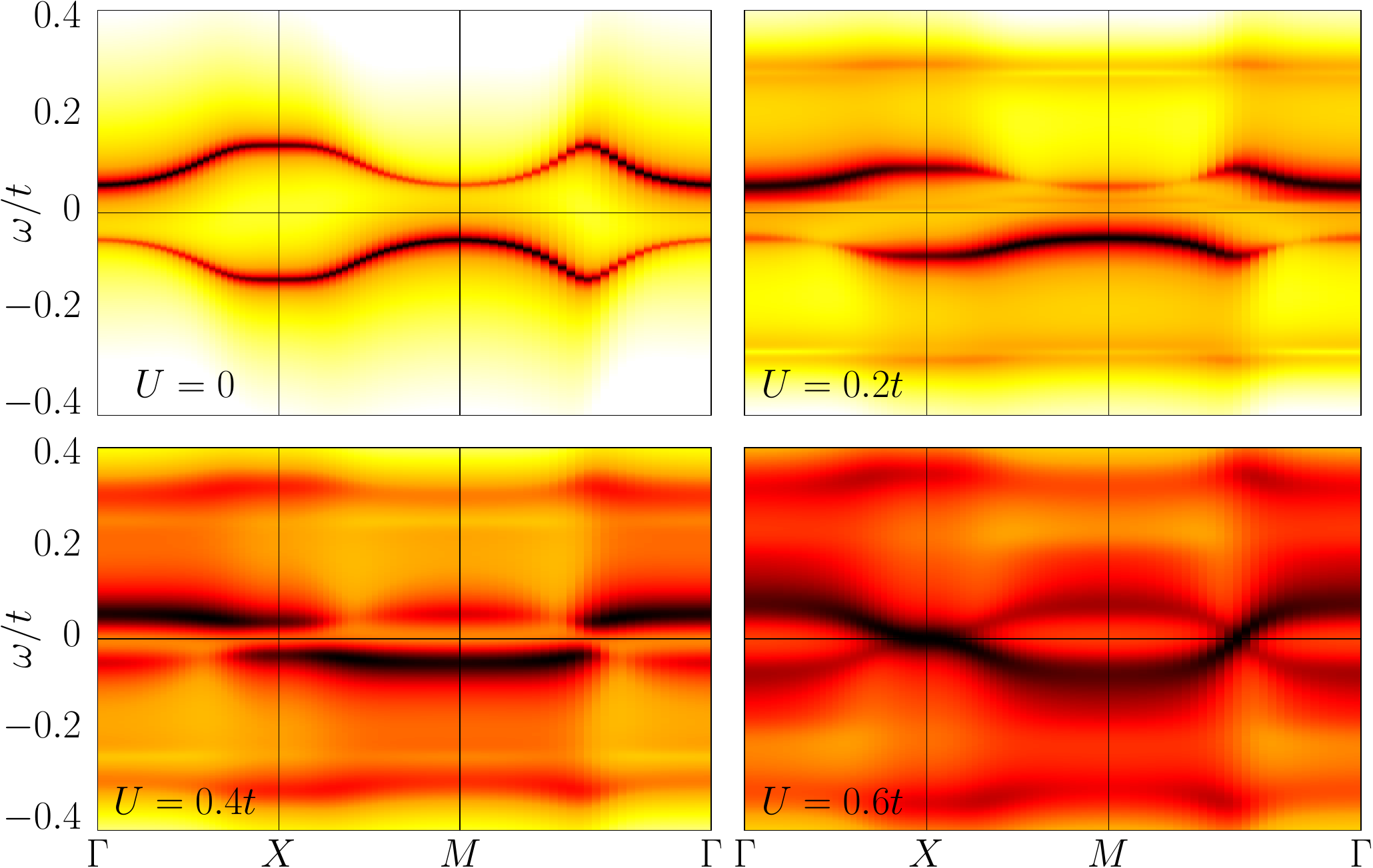}
\caption{Momentum-resolved spectral function $A(\mathbf{k},\omega)$
for the square lattice calculated using IPT for $V=0.5t$, $g=5t$ and $k_BT=0.025t$
for four values of the interaction strength $U=0$ (top left),
$0.2t$ (top right), $0.4t$ (bottom left), and $0.6t$ (bottom right).
The notation on the high symmetry points in the Brillouin zone
follows Fig.~\ref{Fig:nonintBands}(c). 
\label{Fig:bands_g5}}
\end{figure}

To further illustrate the evolution of the bands while approaching
the transition point, we plot in Fig.~\ref{Fig:bands_g5} 
the total momentum-resolved spectral function
$A(\mathbf{k},\omega)=A_d(\mathbf{k},\omega)+A_c(\mathbf{k},\omega)$
calculated using Eq.~\eqref{Eq:DysonPAMloc}
in the gap region along a path in the Brillouin zone described in 
Fig.~\ref{Fig:nonintBands}(c) for four values of the interaction strength. 
The double-peak structure of the bands for vanishing values of $U$ is
connected with the plateaus at $\Gamma$ and $X$ points.
The increasing interaction strength pushes the bands closer to the Fermi energy 
that can be seen on the increase of the value of $\beta G(\beta/2)$ (red line in 
Fig.~\ref{Fig:spec_map_g5}(a)) that shows a maximum at $U\approx0.4t$.
As the bands are moving closer together, they 
simultaneously lose coherence and more spectral weight is pushed to the 
side bands by the increasing effect of the self-energy $\Sigma_d$.
As $\beta G(\beta/2)$ measures the integrated spectral weight in an interval 
around the Fermi energy, its maximum does not match the maximum of 
the spectral function at the Fermi energy $A(0)$ (blue line in 
Fig.~\ref{Fig:spec_map_g5}(a)), although they should become
more similar with decreasing temperature and eventually coincide at $T=0$.

Fig.~\ref{Fig:bands_g5} also illustrates how the increasing self-energy 
is responsible for the formation of the side bands. Their evolution is 
similar to the formation of the Hubbard bands of the SIAM, although
their position deviates for weak interaction strength from the $\omega=\pm U/2$
guess that comes from the atomic limit of the impurity model.
It is more plausible these bands are, in fact, connected with the 
second pair of ABS that emerge in the SCIAM with doublet ground state
and their origin is similar to the origin of the outer peaks in the spectrum
of a single-level impurity that is simultaneously connected to both superconducting 
and metallic baths~\cite{Zalom-2021}.

\subsection{Average CT-HYB expansion order\label{SSec:ExpOrder}}

The average expansion order $\langle k\rangle$ of a CT-HYB simulation
of SCPAM bears the information about the coupling between
the conduction and the impurity band~\cite{Gull-2011} as it can be identified 
with the hybridization energy scaled by inverse temperature,
$\langle k\rangle=\beta\langle \mathcal{H}_{\mathrm{hyb}}\rangle$ where 
$\mathcal{H}_{\mathrm{hyb}}$ is given by Eq.~\eqref{Eq:Hhyb}. 
Therefore, for momentum-independent hybridization $V$ it can be
related to the parameter
$n_{dc}=\frac{1}{N}\sum_{\mathbf{k}}
\langle d_{\mathbf{k}\sigma}^\dag c_{\mathbf{k}\sigma}^{\phantom{\dag}}\rangle$ 
defined in Eq.~\eqref{Eq:densities}, 
$\langle k\rangle=2\beta Vn_{dc}$ (in contrast to the DMFT solution
for the Hubbard model where $\langle k\rangle$ is connected with the
kinetic energy of electrons~\cite{Haule-2007}). The statistics
of $k$ can be accumulated during the CT-HYB simulation and can be used 
to quickly identify the approximate position of the phase boundaries
without the need to measure the Green function or the expectation
value of any operator~\cite{Pokorny-2021}. 

\begin{figure}[ht]
\includegraphics[width=\columnwidth]{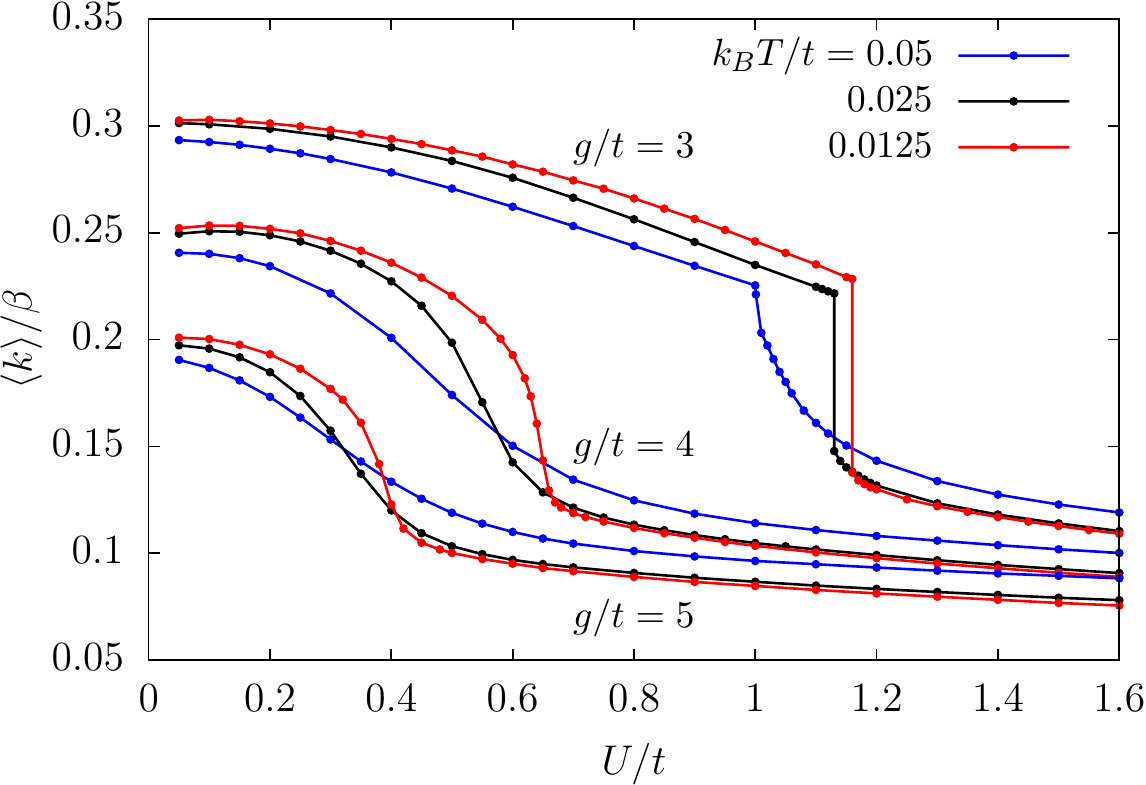}
\caption{Average expansion order $\langle k\rangle$ 
of the CT-HYB simulation scaled by inverse temperature 
$\beta=1/k_BT$ as function of the interaction
strength $U$ for the square lattice, $V=0.5t$, three values of the 
coupling strength $g$ and three temperatures $k_BT=0.05t$ (blue), 
$0.025t$ (black) and $0.0125t$ (red).
\label{Fig:Tdep_pert}}
\end{figure}

In Fig.~\ref{Fig:Tdep_pert} we plotted the scaled average expansion order 
$\langle k\rangle/\beta$ of the CT-HYB simulation as 
a function of the interaction strength $U$ for three values 
of the coupling strength $g$ and
three temperatures $k_BT=0.05t$ (blue), $0.025t$ (black) 
and $0.0125t$ (red). We can use these data to assess the 
effect of the temperature on the position of the phase boundaries.
For $g=3t$ this quantity exhibits a jump at the transition
point between the KI and $\mathrm{SC_-}$ phases. The transition moves
to higher values of $U$ and the jump becomes larger with
decreasing temperature. For $g=4t$ and $5t$ the average 
expansion order exhibits an increasingly abrupt change with 
decreasing temperature around the crossover between the $\mathrm{SC_+}$ 
and $\mathrm{SC_-}$ phases and this crossover again moves 
to larger values of $U$.

For the SCIAM, the 
crossing of the lines for different low-enough temperatures 
$k_BT\ll\Delta$ marks the position of the phase transition at zero 
temperature. This can be proven by mapping the SCIAM in the vicinity
of a critical point to a simple discrete two-level system~\cite{Kadlecova-2019}.
In our case, the low-energy spectrum of SCPAM is continuous and the mapping is
only approximate therefore the lines for different temperatures will not 
cross exactly at the same point, however the crossing of the lines for $k_BT=0.025$ (black)
and $0.0125$ (red) should still give a reasonable guess for the position of the
transition at zero temperatures. A comparison to the NRG data would be needed to assess
the reliability of this guess for lattice models.

\section{Conclusions \label{Sec:Conclusions}}
We studied the properties of a heterostructure consisting of a 
periodic lattice of impurities deposited on
the surface of a BCS superconductor that can be described by SCPAM.
We solved the model within the DMFT framework at half filling using CT-HYB and IPT 
as impurity solvers. CT-HYB provides numerically exact results that showed that
apart from the KI phase there are two superconducting phases we marked $\mathrm{SC_+}$
and $\mathrm{SC_-}$ separated by a crossover
at finite temperatures. These phases represent analogies of the 
$0$ and $\pi$ phases of the SCIAM and were previously identified in 
Ref.~\cite{Luitz-2010} in a model with fixed $\Delta$ but never studied in detail. 
The relation between the phase transitions in the impurity model and in the 
lattice model bound together by the DMFT equations is still an open question~\cite{Bulla-2005}
and here we provide an example of such scenario.
We present phase diagrams at constant temperature that show the evolution of 
the phase boundaries with regard
to the attractive interaction $g$ in the conduction band and the 
repulsive interaction $U$ in the impurity band. At small values of $U$
the interplay between the two interaction strengths leads to a reentrant behavior
of the phase boundary between the KI and the $\mathrm{SC_+}$ superconductor.
For larger values, the Coulomb interaction favors the $\mathrm{SC_-}$ superconducting phase 
over the KI by lowering the critical value of the attractive interaction strength. 
A similar effect was described in an experimental setup in which the presence of a spin-1/2 transition 
metal phtalocyanine molecules on the surface of a two-dimensional superconductor 
enhances the superconducting pairing~\cite{Yoshizawa-2017}. As the CT-HYB 
calculation is performed in imaginary time, there is no direct way to access 
the spectral functions except performing an ill-defined analytic continuation.
Therefore we used the approximate IPT method that provides
reliable results for the square lattice. It correctly describes the KI and $\mathrm{SC_+}$
phases as well as crossover between the $\mathrm{SC_+}$ and $\mathrm{SC_-}$ phases. The in-gap 
bands follow a crossing-like scenario in which the spectral weight is transferred 
to the Fermi energy in the crossover region. Furthermore, a second pair of in-gap peaks 
is formed at higher energies. Unfortunately, the current
implementation of IPT fails to describe these effects for the triangular lattice.
Its reliability in the $\mathrm{SC_-}$ phase is also questionable due to 
the inherent failure of the underlying 2PT solver to correctly describe a spinful 
ground state of the auxiliary impurity model. 
In the last part we discussed the temperature dependence of the 
average expansion order of the CT-HYB algorithm.
This quantity can be calculated very effectively and bears the information
about the change of the phase boundaries with regard to the temperature.

\begin{acknowledgments}
We acknowledge fruitful discussions with V. Jani\v{s}, T. Novotn\'y and M. \v{Z}onda.
This research was supported by the Czech Ministry of Education, Youth and Sports
program INTER-COST, grant No. LTC19045 and through the project e-INFRA CZ (ID:90140).
\end{acknowledgments}

\appendix
\section{LDOS for square and triangular lattices\label{App:DOSzero}}
\begin{figure}[ht]
\includegraphics[width=\columnwidth]{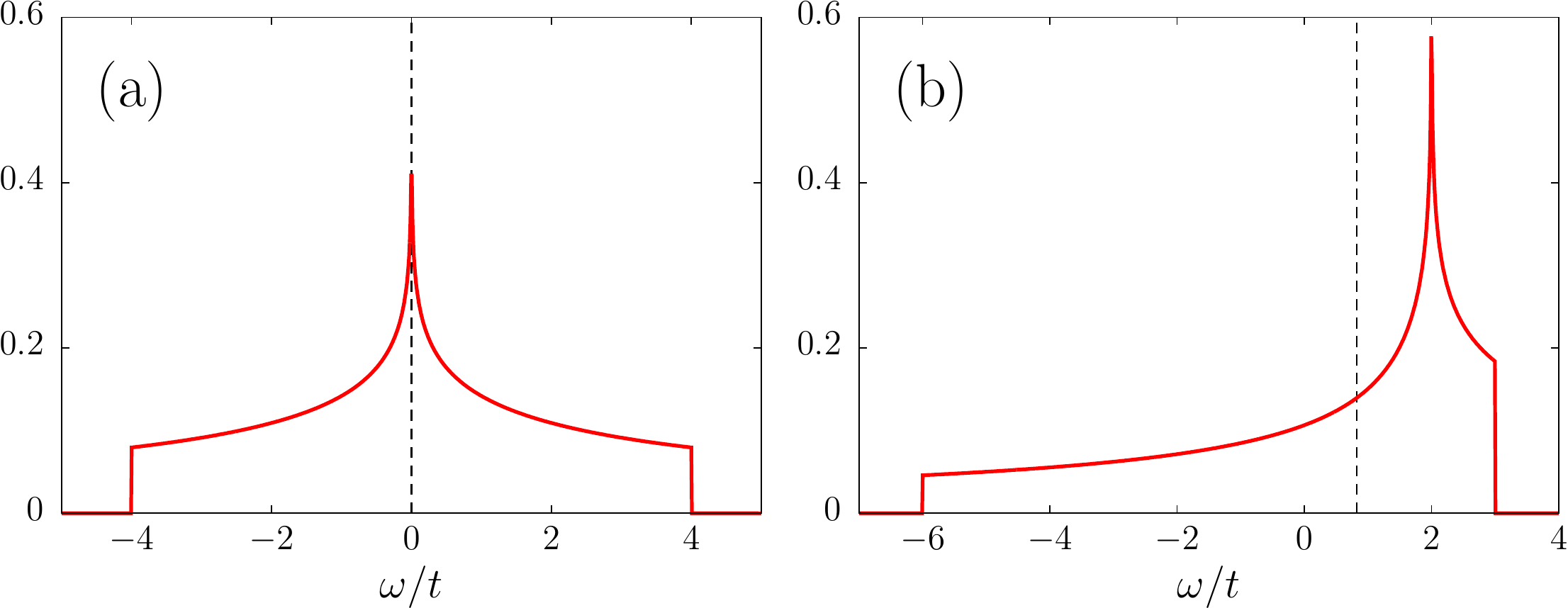}
\caption{The non-interacting LDOS for (a) square lattice, Eq.~\eqref{Eq:DOSsquare}
and (b) triangular lattice, Eq.~\eqref{Eq:DOStri}. 
Dashed lines mark the position of the chemical potential that corresponds to half filling.
\label{Fig:nonintDOS}}
\end{figure}

The momentum summation in Eq.~\eqref{Eq:Gloc} can be calculated effectively
using the Hilbert transform and the non-interacting LDOS for the given 
lattice which can be calculated from the local Green function.
For the square lattice with dispersion given by~Eq.~\eqref{Eq:dispSq}
it reads~\cite{Kogan-2020}
\begin{equation}
G_{\mathrm{loc}\square}(z)
=\frac{1}{(2\pi)^2}\int\frac{d\mathbf{k}}{z-\varepsilon_{\mathbf{k}\square}}
=\frac{2}{\pi z}K\left(\frac{16t^2}{z^2}\right),
\end{equation}
where the integration is over the first Brillouin zone and
\begin{equation}
K(m)=\int_0^{\pi/2}\!\!\!\!\frac{d\theta}{\sqrt{1-m\sin^2(\theta)}}
\end{equation}
is the complete elliptic integral of the first kind. Using the identity
\begin{equation}
K(m)=\frac{1}{\sqrt{m}}\left[K\left(\frac{1}{m}\right)-isK\left(1-\frac{1}{m}\right)\right],
\end{equation}
where $s=\sgn (\Imm m)$ we obtain the LDOS
$A_\square(\omega)=-\Imm G_{\mathrm{loc}\square}(\omega+i0)/\pi$ that reads
\begin{equation}
\label{Eq:DOSsquare}
A_\square(\omega)=\frac{1}{2\pi^2t}K\left(1-\frac{\omega^2}{16t^2}\right)\Theta(16t^2-\omega^2).
\end{equation}
For the triangular lattice with dispersion given by Eq.~\eqref{Eq:dispTr} 
we obtain~\cite{Kogan-2020}
\begin{equation}
G_{\mathrm{loc}\triangle}(z)=\frac{1}{\pi\sqrt{z_0}}K\left(\frac{4r}{z_0}\right),
\end{equation}
where $r=\sqrt{3-z/t}$ and $z_0=(r+3)(r-1)^3/4$. The analytic continuation
to the real axis must be done carefully due to the complicated structure of the 
argument. The LDOS reads
\begin{equation}
\label{Eq:DOStri}
A_\triangle(\omega)=
\begin{cases}
\frac{1}{4\pi\sqrt{z_1}}K\left(\frac{z_2}{z_1}\right)&\mbox{~for~}-6t<\omega<2t, \\
\frac{1}{4\pi\sqrt{z_2}}K\left(\frac{z_1}{z_2}\right)&\mbox{~for~}2t<\omega<3t, \\
0&\mbox{otherwise},
\end{cases}
\end{equation}
where $z_1=4q$, $z_2=(3-q)(1+q)^3/4$, and $q=\sqrt{3-\omega/t}$.
The LDOS for the square and triangular lattices are plotted in Fig.~\ref{Fig:nonintDOS}.

\section{Non-interacting band structure\label{App:Nonint}}
\begin{figure}[ht]
\includegraphics[width=\columnwidth]{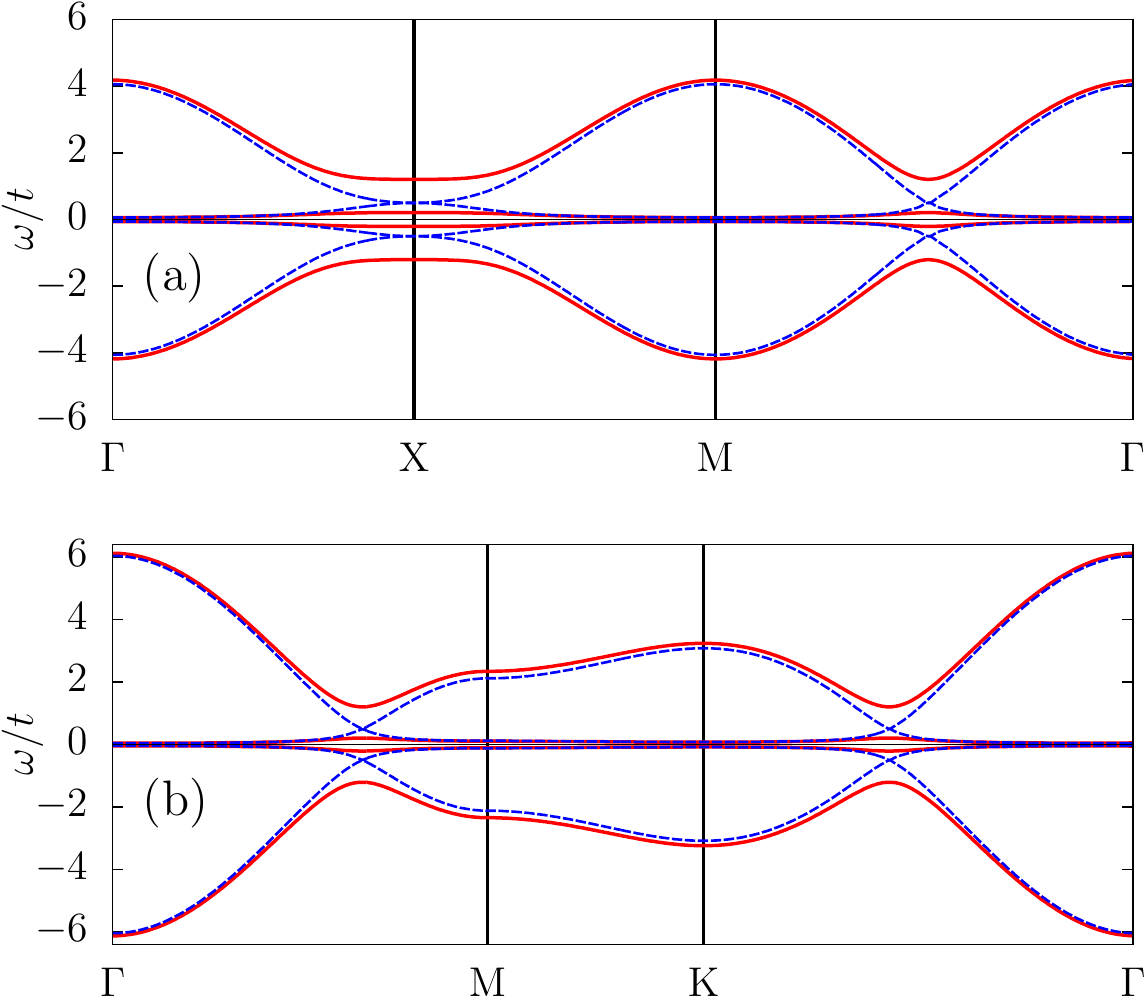}\\
\vspace{4mm}
\includegraphics[width=0.9\columnwidth]{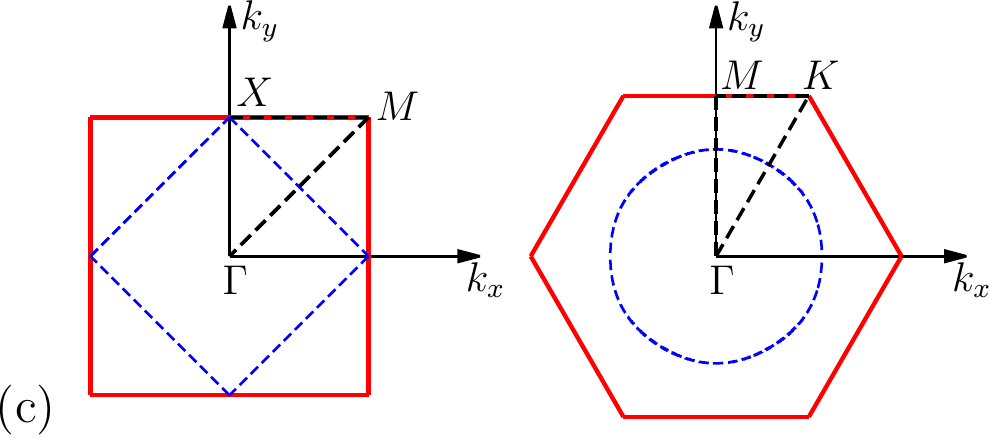}
\caption{Non-interacting ($U=0$) band structure of the SCPAM for 
(a) square and (b) triangular lattice for $V=0.5t$ and $\Delta=0$
(blue dashed line), and $\Delta=t$ (red solid line).
(c) First Brillouin zones for square (left) and triangular (right) 
lattice. Blue dashed line is the Fermi surface for $\Delta=V=0$. Black dashed
line marks the path along which the data in (a) and (b) and in 
Fig~\ref{Fig:bands_g5} are plotted.
\label{Fig:nonintBands}}
\end{figure}

The $U=0$ band structure of the SCPAM can be calculated from
the Green function in Eq.~\eqref{Eq:GF0}.
The determinant 
$D(\mathbf{k},\omega)=\mathrm{Det}[G^{-1}_0(\mathbf{k},\omega)]$ 
reads
\begin{equation}
\begin{aligned}
D(\mathbf{k},\omega)&=
(\omega^2-\varepsilon^2)(\omega^2-\zeta_\mathbf{k}^2) \\
&-2|V|^2(\omega^2+\varepsilon\varepsilon_{\mathbf{k}})+|V|^4,
\end{aligned}
\end{equation}
where $\zeta^2_\mathbf{k}=\Delta^2+\varepsilon^2_\mathbf{k}$.
Its zeros $\omega(\mathbf{k})$ mark the positions of the poles of the Green function 
and read 
\begin{equation}
\begin{aligned}
\omega(\mathbf{k})=&
\pm\frac{1}{\sqrt{2}}\Bigg[2|V|^2+\zeta^2_\mathbf{k}+\varepsilon^2 \nonumber \\
&\pm\sqrt{(\zeta_\mathbf{k}^2-\varepsilon^2)^2+4|V|^2(\zeta^2_\mathbf{k}+
\varepsilon^2+\varepsilon\varepsilon_\mathbf{k})}\Bigg]^{1/2}.
\end{aligned}
\end{equation}
For $V=0$ the band structure
simplifies to $\omega(\mathbf{k})=\{\pm\varepsilon,\pm\zeta_\mathbf{k}\}$.
In Fig.~\ref{Fig:nonintBands} we plot the bands for $U=0$
for square, Fig.~\ref{Fig:nonintBands}(a), and triangular, Fig.~\ref{Fig:nonintBands}(b),
lattice along the selected path through the Brillouin zone (black dashed line in Fig.~\ref{Fig:nonintBands}(c)). 
For $\Delta=0$ (dashed blue lines) the model shows a very narrow hybridization gap at 
the Fermi energy. For $\Delta>0$ an additional gap opens between the conduction
and the impurity bands.

\section{Effect of hybridization and temperature on the 
non-interacting model\label{App:Nonint2}}
\begin{figure}[htb]
\includegraphics[width=\columnwidth]{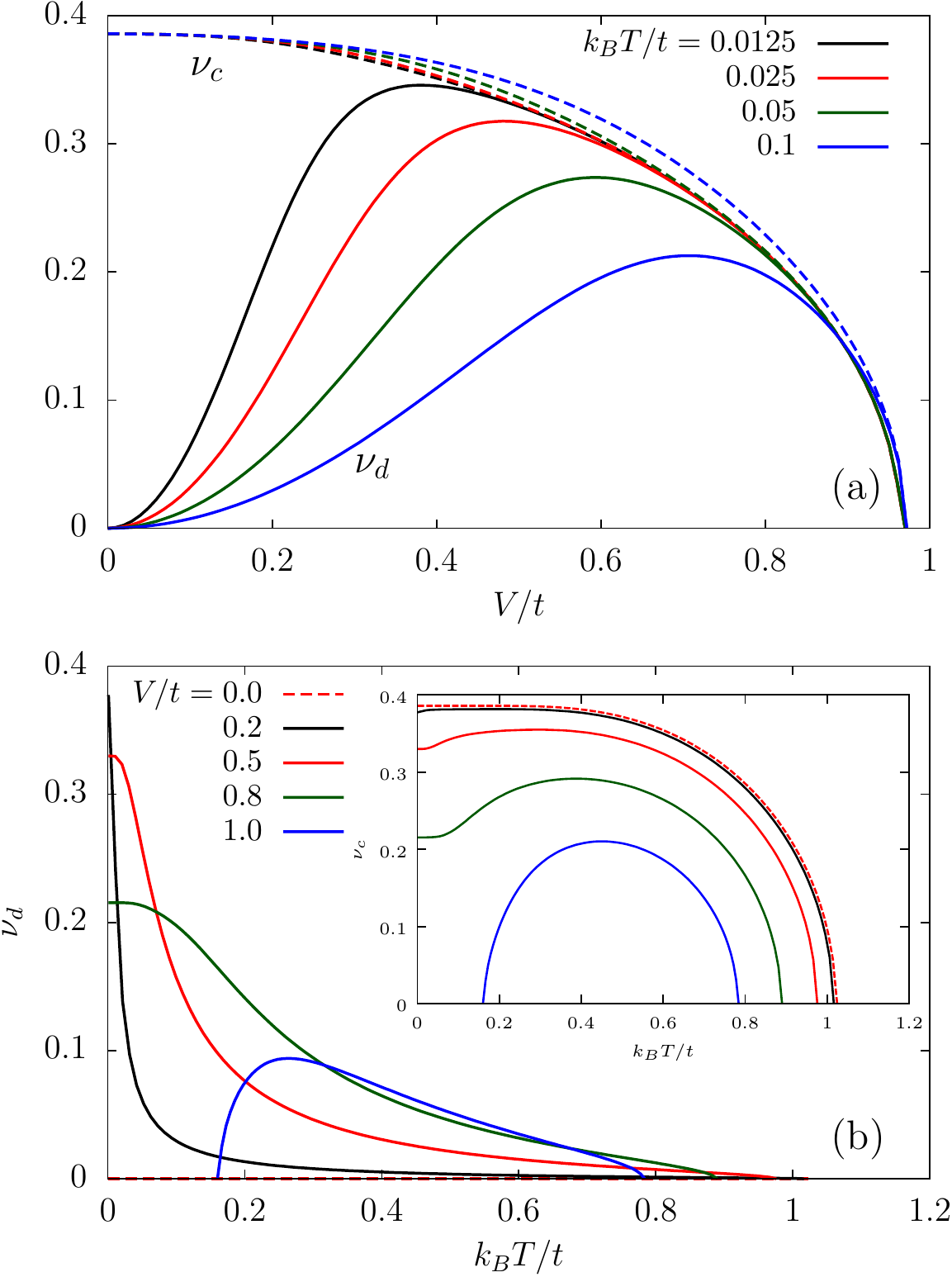}
\caption{(a) Induced pairing $\nu_d$ (solid lines) and intrinsic pairing $\nu_c$ 
(dashed lines) as functions of the hybridization $V$ for selected temperatures
at $U=0$ and $g=5t$.
(b) Induced pairing $\nu_d$ (main panel) and intrinsic pairing $\nu_c$ (inset)
as functions of the temperature for selected values of the hybridization
at $U=0$ and $g=5t$.
\label{Fig:TdepNonint}}
\end{figure}

All results in Sec.~\ref{Sec:Results} (except Fig.~\ref{Fig:Tdep_pert}) were calculated for the
fixed value of the hybridization $V=0.5t$ and temperature $k_BT=0.025t$.
Here we briefly discuss the effect of these parameters on the non-interacting 
($U=0$) model. In Fig.~\ref{Fig:TdepNonint}(a) we plot the dependence
of the induced pairing $\nu_d$ (solid lines) and intrinsic pairing $\nu_c$ 
(dashed lines) as functions of $V$ for several temperatures
and $g=5t$. The increasing hybridization weakens the superconducting correlations 
in the conduction band measured by $\nu_c$ by strengthening the pair breaking 
effect of the magnetic impurities. The effect of the hybridization on the induced 
pairing $\nu_d$ is more complicated, as for small values of $V$ it promotes
the proximity effect leading to the increase of the pairing in the impurity band, while
for larger values the induced pairing decreases at the same rate as the intrinsic 
pairing in the conduction band. The maximum of $\nu_d$ for $k_BT=0.025t$ lies at 
$V\approx0.48t$, close to the value for which the results in the main text are plotted.

In Fig.~\ref{Fig:TdepNonint}(b) we plot the induced pairing $\nu_d$ (main panel)
and the intrinsic pairing $\nu_c$ (inset) for $g=5t$ and $U=0$ as functions of 
the temperature. For $T=0$ the values of $\nu_d$ and $\nu_c$ coincide for any $V>0$.
For vanishing values of $V$ the intrinsic pairing decreases 
with increasing temperature according to the standard BCS theory and vanishes 
at the critical value $T_c$, followed by the induced pairing. For larger values of 
$V$ the dependence is non-monotonic and for $V=t$ (blue line) we observe 
the reentrant behavior of the superconductivity. This effect is also discussed 
in Ref.~\cite{Oei-2020} and our result is in agreement with their 
conclusions that the reentrant behavior is not an effect of the electron correlations,
as we observe it already at $U=0$, but rather a result of the subtle interplay 
between the electron tunneling and superconducting pairing.

\newpage


%

\end{document}